\documentclass[showpacs,manuscript]{revtex4-1}

\usepackage{graphicx}
\usepackage{amssymb}
\usepackage{color}
\usepackage{tabularx}
\usepackage{multirow}
\usepackage[dvips]{epsfig}

\begin{document}

\newcommand{\be}{\begin{equation}}
\newcommand{\ee}{\end{equation}}
\newcommand{\bea}{\begin{eqnarray}}
\newcommand{\eea}{\end{eqnarray}}
\newcommand{\Sv}{\hat{{\bf S}}}

\title{Structural Reorganization of Parallel Actin Bundles by Crosslinking Proteins:  Incommensurate States of Twist}
\author{Homin Shin and Gregory M. Grason}
\affiliation{Department of Polymer Science and Engineering, University of Massachusetts, Amherst, MA 01003}
\begin{abstract}
We construct a coarse-grained model of parallel actin bundles crosslinked by compact, globular bundling proteins, such as fascin and espin, necessary components of filapodial and mechanosensory bundles.  Consistent with structural observations of bundles, we find that the optimal geometry for crosslinking is {\it overtwisted}, requiring a coherent structural change of the helical geometry of the filaments.  We study the linker-dependent thermodynamic transition of bundled actin filaments from their native state to the overtwisted state and map out the ``twist-state'' phase diagram in terms of the availability as well as the flexibility of crosslinker proteins.  We predict that the transition from the uncrosslinked to fully-crosslinked state is highly sensitive to linker flexibility:  flexible crosslinking smoothly distorts the twist-state of bundled filaments, while rigidly crosslinked bundles undergo a phase transition, rapidly overtwisting filaments over a narrow range of free crosslinker concentrations.  Additionally, we predict a rich spectrum of intermediate structures, composed of alternating domains of sparsely-bound (untwisted) and strongly-bound (overtwisted) filaments.  This model reveals that subtle differences in crosslinking agents themselves modify not only the detailed structure of parallel actin bundles, but also the thermodynamic pathway by which they form.
\end{abstract}

\maketitle

\section{Introduction}
Parallel actin bundles are highly organized structures crucial to diverse range of cellular function, from mechanosensory specializations such as microvilli, stereocilia and neurosensory bristles to the highly dynamic filapodial protrusions of cell cytoskeletons~\cite{pollard, revenu}.  These assemblies share a common structural organization:  axially-aligned actin filaments of uniform polarity, densely-assembled into an ordered hexagonal array and interspersed with a crosslinking array of {\it actin bundling proteins}.  Multiple bundling proteins have been identified from parallel actin bundles {\it in vivo}, though the type and composition bundling proteins varies significantly between different cell types~\cite{bartles}.  Primary examples of actin bundling proteins, fascin and espin, are known to be integral components of filapodia~\cite{vignjevic, faix} and stereocilia bundles~\cite{bartles_zheng}, respectively.  It is believed that the array of multiple bundling proteins affords cells the ability to form actin bundles with variable properties, such as size~\cite{loomis, claessens_pnas_08} and rigidity~\cite{shin_mahadevan, claessens_nat_06, howard_08}, though little is understood about how distinct features of bundling proteins specifically modify the assembly of actin filaments into bundles.  

Structural studies of {\it in vitro}~bundles \cite{derosier_censullo, tilney_derosier_mulroy, derosier_tilney, angelini, purdy, claessens_pnas_08, shin} suggest that a key aspect of the formation of parallel bundles is the ability of crosslinking proteins to modify the twist of actin filaments.  In bundles, the helical symmetry of constituent filaments is modified from its native -13/6 geometry:  a left-handed helix rotating through 6 turns per 13 monomer repeat.  This native geometry is poorly suited for the hexagonal symmetry of the array, which favors co-registry of crosslinked monomers on neighbor filaments~\cite{derosier_tilney}.  Electron diffraction studies of fascin-crosslinked bundles reveal that filaments are overtwisted to a $-28/13$ symmetry, corresponding to a change of roughly - 0.01 monomers/turn, a distortion which is consistent with more recent observations of espin-mediated bundles~\cite{purdy, shin}.  Despite the apparently similar structural change induced by crosslinking, fascin- and espin-mediated bundles exhibit a dramatically different sensitivity to concentration of available crosslinkers in {\it in vitro} systems.  Based on small-angle x-ray studies, Claessens {\it et al.} found that overtwist of filaments in fascin-mediated bundles is sensitive to the concentration of available crosslinker, with helical filament symmetry varying continuously from {\it native} to {\it fully overtwisted} symmetry~\cite{claessens_pnas_08}.  In comparison, a recent study of espin cross-linked bundles found that above a critical concentration of crosslinker, bundles lock into the fully overtwisted state, with little or no further sensitivity to espin concentration~\cite{shin}.  Evidently these two compact, globular bundling proteins, fascin and espin, are capable of forming bundles of the apparently same overtwisted structure, though in each case the fully-bundled state is approached via a different pathway of states at intermediate crosslinker concentrations.   

This rich phenomenology raises a number of questions about the role of filament twist in the assembly mechanism of protein-mediated bundles.  What is the mechanical cost associated with distorting bundled filaments from their native geometry, and how does this cost effect the bundling transition?  What is the role of the torsional rigidity of the crosslinking bonds themselves?  Most puzzling, what is the nature of the states of {\it intermediate} twist observed for fascin-mediated bundles?  

In this article, we analyze a quantitative model that describes the complex interplay between the optimal geometry required by crosslinking in actin in hexagonal bundles and the cost of distorting filaments from their ideal helical symmetry.  Based on a lattice model proposed in ref. \cite{shin}, we identify a unique crosslinked-bundle groundstate with -28/13 symmetry that allows for an optimal number of ``ideally" oriented crosslinking bonds.  We study the thermodynamic transition from untwisted, unbound filaments to fully-bound, overtwisted bundles driven by increasing the concentration, or chemical potential, of available crosslinkers.  A coarse-grained model of parallel bundles, maps the linker-induced overtwist of actin filaments onto a {\it commensurate-incommensurate} phase transition.  We find that this transition takes place by a surprisingly complex coherent restructuring of filaments in the bundle, in which bundles possess localized bands, or domains, of {\it native} (-13/6) and {\it overtwisted} (-28/13) filaments.  The overtwisting of bundled filaments then proceeds as the fraction of overtwisted bundles increases continuously from 0, in the absence of crosslinkers, to 1 in excess of available crosslinkers.  Owing to the fundamental role of {\it elastic} distortions in this model, the bundling transition is found to be extremely sensitive to the flexibility of the crosslinking bounds.  For sufficiently rigid crosslinkers, bundles pass to the overtwisted state via a sharp, thermodynamic transition; while for bundles held together by relatively flexible linkers, a smooth transition to a maximum state of twist is predicted.  The primary conclusion of this study is that differences in the sensitivity of the bundling transition to crosslinker concentration observed from fascin- and espin-mediated bundles derive from distinctions of the flexibility of crosslinker to actin bonds.  Hence, we have identified the flexibility of the crosslinking bonds provided by bundling proteins as a key parameter controlling not only the structure, but the process, by which parallel actin bundles are formed in different cell types.  

\section{Lattice Model of Crosslinking in Parallel Actin Bundles}
\begin{figure}
\centering
\center \epsfig{file= 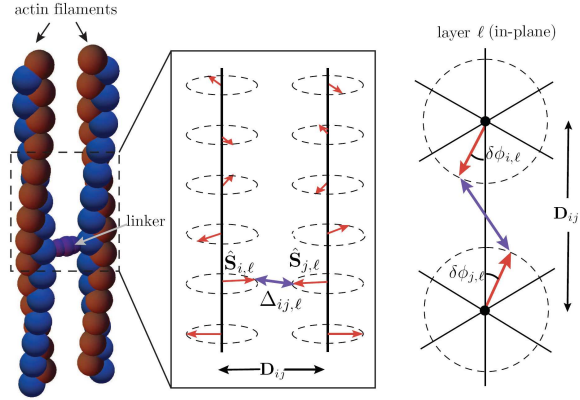, width=3.25in}
\caption{A schematic picture of two actin filaments crosslinked by a bundling proteins (purple) and its lattice model representation in a box, where  G-actin monomers are described by a set of vectors, $\Sv_{ i, \ell}$.  The top view of layer $\ell$ is also shown with the angular deviation $\delta \phi_{i, \ell}$ of the monomer from the lattice direction ${\bf D}_{ij}$. }
\label{fig: spinlattice}
\end{figure}

We model a bundle as a parallel, hexagonally-ordered array of actin filaments with fixed nearest neighbor spacing $D \approx 17$ nm, consistent with structural observations~\cite{claessens_pnas_08, purdy}.  The positions of the monomeric, G-actin are described by a set of vectors, $a \Sv_{ i, \ell}$, that point from the center line of the $i$th actin filament of the lattice to the center of the $\ell$th monomer {\it along} the filament, where $a \simeq~3.75$ nm roughly the diameter of G-actin (see Fig. \ref{fig: spinlattice}).  In the native twist state, these vectors precess around the centerline of filaments at a constant angular rate of $\omega_0 =  12 \pi /13$ per monomer (i.e. 6 rotations per 13 monomer repeat)~\cite{holmes}.   We describe torsional distortions with the following elastic energy~\cite{footnote},
\begin{equation}
\label{eq: twist}
E_{twist} = \frac{C}{2} \sum_{\ell, i} (\Delta \phi_{i, \ell} - \omega_0)^2 ,
\end{equation}
where $C$ is the torsional elastic modulus of actin filaments~\cite{tsuda}, and $\Delta \phi_{i, \ell} =\phi_{i, \ell+1} - \phi_{i, \ell}$ is the rotation angle between successive monomers along the $i$th filament, with $\phi_{i, \ell}$ the angle $\Sv_{ i, \ell}$ of the $\ell$th monomer direction in the plane of lattice order.

In our model, crosslinking between neighboring filaments in the bundle occurs between pairs of monomers at the same vertical layer, $\ell$, shown schematically in Fig. \ref{fig: spinlattice} .  Because globular bundling proteins like epsin and fascin have compact size, $\sim 5-7~{\rm nm}$ in diameter~\cite{bartles_zheng, sedeh}, in comparison the lattice between filament in the bundle $\sim 15-20~{\rm nm}$, crosslinking occurs preferentially when monomers on adjacent filaments are minimally separated from one another.  Yet, due to incommensurate helical symmetry of actin filaments, bundles are forced to accomodate crosslinks with some degree of ``misfit" between crosslinked monomers.  We describe this effect with the following simple elastic model for enthalpy of crosslinking between monomers,
\begin{equation}
\label{eq: binding}
E_{binding} = \sum_{\ell, \langle ij \rangle} n_{ij, \ell} \Big[-\epsilon_b + \frac{U}{2}( \delta \phi_{i, \ell}^2 +  \delta \phi_{j, \ell}^2 ) \Big] ,
\end{equation}
where the sum is over neighbor filaments, $\langle i j \rangle$. In this formula, $n_{ij, \ell}$ is 1 if the bond between $i$ and $j$ at $\ell$ is occupied, and 0 if empty, $\epsilon_b$ describes the energy gain of a perfectly-aligned, crosslinking bond between 2 monomers, and the final term in the parentheses describes the energetic cost of distorting the bond from its ideal geometry.  Here, $\delta \phi_{i, \ell}$ is the angular deviation between $\Sv_{i, \ell}$ and ${\bf D}_{ij}$, the lattice vector separating $i$ and $j$.  Finally, $U$ is a parameter describing the ``elastic cost" of distorting the aligned bond between monomers.  For example, if this cost could be described purely in terms of a simple linear spring energy, $k_b(\Delta_{ij, \ell}-\Delta_0)^2/2$, which penalizes changes in length, $\Delta_{ij, \ell}$, of the monomer-monomer separation from a zero stretch length, $\Delta_0 = D-2a$ the equilibrium size of crosslinks,  the effective elastic parameter in (\ref{eq: binding}) becomes $ U \simeq 2 k_b a^2$.  

The bundling of actin is sensitive to the concentration of free crosslinkers in solution.  We therefore study the thermodynamics of crosslinking at a fixed chemical potential, $\mu$.  This accounts for the equilibrium free energy cost of removing a free crosslinking protein from solution and adding it to a bundle, and therefore, $\mu$ is related to the free crosslinker concentration by $c_{free} \propto e^{\mu/k_B T}$.   

\section{Ideal Crosslinking Geometry in Bundles}

To describe the overtwist transition of parallel bundles, it is necessary to understand the optimal geometry of highly crosslinked bundles, as well as the low energy pathways to this state from the native actin geometry.  First, we briefly demonstrate the structure of optimally-packed actin bundles in terms of geometric considerations (see supporting materials for full details).  The model described above is highly frustrated, a generic feature of hexagonally-organized  filament assemblies, well-studied in the context of counterion mediated biopolymer bundles~\cite{angelini, angelini_epje, grason_bruinsma_prl_06, grason_bruinsma_pre} as well as helically-ordered phases of DNA~\cite{harreis, kornyshev_rmp, grason_epl}.    Here, we consider the configurations for which crosslinking bonds are {\it perfectly aligned} to the lattice directions (that is, $\delta \phi_{i, \ell} =0$ for all $n_{ij, \ell} =1$ ) and for which configurations the number of perfectly oriented bonds is {\it maximal}.    

Perfectly aligned configurations require a subset of the actin monomers to align with a sixfold lattice direction.  We construct structures which have alternating sequences of sections of $- 6 m / n$ symmetry -- each successive monomer is rotated by $- 2 \pi n / (6 m)$.  Here, $n$ and $m$ are integers so that $m$th monomer lines up with lattice direction of $2\pi n/6$, allowing for a perfectly aligned crosslinking bond to form.  It is not difficult to show that among these commensurate helical geometries the $-24/11$ and $-30/14$ structures are particularly close the native geometry of actin, differing only by $0.69\%$ and $1.1\%$, respectively, in terms rotation angle per monomer.  This proximity to the native geometries confers upon them an especially low {\it twist} cost among all possible ideal crosslinking states.  To determine the bundle structure with the maximum number of bonds, we therefore considered periodic states with a composite symmetry, possessing $N_4$ numbers of 4-monomer sections with -24/11 symmetry and $N_5$ numbers of 5-monomer sections with -30/14 symmetry.  To construct crosslinks, it is not sufficient to consider aligned monomers to lattice directions from a single filament, as crosslinking requires the {\it co-orientation} of monomers on neighbor filaments at the same vertical layer along the filaments.  Hence, it is necessary to consider the three-dimensional geometry of possible multi-filament structures arrayed on the hexagonal lattice.  

\begin{figure}
\centering
\includegraphics[angle=0.0, width=3.35in]{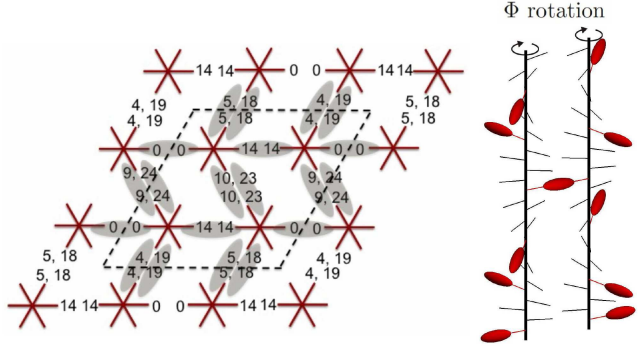}
\caption{The unit cell of maximally crosslinked-bundle groundstate in the plane of hexagonal order.   The repeat unit contains a single filament geometry translated and rotated vertically to obtain the geometry of the 4 filaments.  Numbers represent the vertical layers of co-oriented monomers with the lattice direction, and gray bars, crosslinkers.  On the right, the 28-monomer vertical repeat geometry of 2 neighbor filaments, highlighting definition of $\Phi$, the coherent rotation of filaments from the ideal crosslinking geometry. }
\label{perfectpacking}
\end{figure}

Based on an extensive numerical search of composite  $-24/11$ and $-30/14$ structures up to 102 monomers per repeat length, we find a maximum crosslinking density for $N_4=2$ and $N_5=4$, which has 6 crosslinks along every 28 monomer length of actin filament (see Fig.~\ref{perfectpacking}).  Notice that this composite structure has $-4N_4(11/24) - 5N_5 (14/30) = -13$ net turns per 28-monomer repeat.  The overall symmetry and bond/monomer stoichometry of this ideal geometry are in perfect agreement with careful structural studies of overtwist actin bundles, formed by both espin and fascin crosslinkers, which also have an overtwisted $-28/13$ structure and crosslinks spaced at 4- and 5-monomer intervals along filaments~\cite{derosier_tilney}. Though it has non-hexagonal symmetry, the composite filament-bond structure of this perfect packing geometry can be repeated to construct a parallel bundle of arbitrary size.  This unique crosslinker geometry serves as the overtwist groundstate of our model.

\section{Untwisting Overtwisted Bundles:  Coarse-Grained Theory}

Having identified the limiting geometry of unbound filaments (13-fold helical symmetry) and fully bound filaments (28-fold helical symmetry), we consider the thermodynamic progression of filament twist as crosslinker density in bundles increases.  Although the detailed structure of the overtwisted -28/13 structure is somewhat complex, the underlying screw-symmetry of actin filaments and the in-plane periodic order imbue this state with a rather simple symmetry under {\it coherent rotations of each filament by $2 \pi/28$ around its axis}.  A $2 \pi/28$ rotation of each filament in the ideal crosslinking configuration in Fig. \ref{perfectpacking} followed by a rearrangement of monomers and bonds within the unit cell recovers an equivalently ideal geometry, with 6 perfectly aligned crosslinks per 28 monomers (see supporting information).  Hence, the ultimate function of the complex pattern of crosslinking bonds is to lock the bundle into a -28/13 twist-symmetry and constrain the azimuthal orientation of this structure to within one of 28-fold bonding free energy minima.  Competing with this tendency is the intrinsic torsional elastic energy of filaments which favors unwinding the overtwisted state to the -13/6 symmetry, making it costly for the bundle to maintain a 28-fold commensurate bond geometry along its length.

To model the free energy gain of crosslinking, we construct actin bundles in the fully-overtwisted state with a net degree of twist, $\Omega_0 =  2 \pi(13 /28)$, and consider the low-energy distortions that {\it untwist} the 28-fold commensurate geometry of bundles as crosslinkers unbind from bundles.  The analysis is based on a coarse-graining of the model described by eqs. (\ref{eq: twist}) and (\ref{eq: binding}).  In particular, we decompose the monomer orientations in terms of two angular deviations from {\it homogenuously overtwisted} filaments:  $\tilde{\phi}_{i, \ell}$, which describes short-lengthscale monomer relaxations within a 28-monomer repeat length, and $\Phi_\ell$, describing the slow, coherent rotations of filaments on much longer length scales.  In terms of the angle a monomer direction makes in the plane of hexagonal order, we define,
\begin{equation}
\phi_{i,\ell} = \Omega_0 \ell + \Phi_\ell +\tilde{\phi}_{i,\ell} ,
\end{equation}
where we restrict the lengthscale reorganization to sum to zero rotation within a 28-monomer repeat length, $\sum_{\ell = \ell_0}^{\ell_0+28} \tilde{\phi}_{i,\ell} = 0$ so that the net rotation of filaments away from the overtwisted state is $\Phi_\ell$.  As a description of the long-lengthscale structure of bundles, $\Phi_\ell$ serves as the order parameter, fully describing the underlying state of our model:  overtwisted states commensurate with the ideal crosslinking geometry of the bundle correspond to $\Phi_\ell = 2 \pi m /28$ for any integer, $m$.  

Due to the separation of length scales between the deformations described by $ \Phi_\ell$ and $\tilde{\phi}_{i,\ell}$ the elastic twist energy approximately decouples these degrees of freedom,
\begin{equation}
\label{eq: twist2}
E_{twist}=\frac{C}{2} \sum_{i, \ell} \big[(\Delta \Phi_\ell - \delta \omega_0)^2 + (\Delta \tilde{\phi}_{i,\ell} )^2 \big] ,
\end{equation}
where $\delta \omega_0 = \omega_0 - \Omega_0 = 2 \pi/364$ is the overtwist distortion per monomer from the native to the homogeneously-twisted -28/13 state and we have implicitly assumed that $\Delta \Phi_\ell$ is approximately constant over the length of a 28-monomer repeat.  

The final step of our coarse-graining minimizes the twist and binding free energy, eqs. (\ref{eq: twist2}) and (\ref{eq: binding}), over the distributions of bonds, $n_{ij, \ell}$, and $\tilde{\phi}_{i,\ell}$, the short length-scale angular reorganizations for a given value of $\Phi$.  We perform this minimization by analyzing a 28-monomer repeat of the ideal bonding configuration Fig. \ref{perfectpacking}, requiring a net rotation of this structure by $\Phi$, and finding the subset of the 6 perfectly aligned crosslinks for which net free energy of binding, including the cost of elastic distortion, is minimal (see Appendix \ref{app: V}).   The result is the binding free energy per monomer, $V(\Phi)$, that depends only on mean value of $\Phi$ within a 28-monomer repeat (the coarse-grained unit of our model),
\begin{equation}
\label{eq: V}
V(\Phi) = {\rm min}_{n_b, m} \Big[-\frac{n_b}{2}  \big(\mu- \mu_c (n_b) \big) + \frac{K_{n_b}}{2} \Big(\Phi+ \frac{2 \pi m}{28} \Big)^2 \Big]/28 ,
\end{equation}
where $n_b$ is the number of crosslinked monomers per 28-monomer repeat, which varies from $n_b=0$ (unbound) to $n_b=6$ (fully bound) and the minimization of $m$ reflects rotational symmetry of the binding free energy, $V(\Phi+2 \pi/28)= V(\Phi)$.   Here, $\mu_c= -\epsilon_b + \delta \mu_{n_b}$, where $\delta \mu_{n_b}$ represents the excess torsional elastic energy per bond needed to distort the homogeneously overtwisted filament into a state where the $n_b$ bonds are perfectly aligned with bond directions.   This torsional cost represents an offset to the binding free energy of the state with $n_b$ monomers proportional to $C$ that increases with number of bonds:  $\delta\mu_0=0$; $\delta \mu_1=0$; $\delta \mu_2=0$; $\delta \mu_3=0.00058 C$; $\delta \mu_4=0.00087C$; $\delta \mu_5=0.00191C$; and $ \delta \mu_6=0.00261C$.  The term proportional to $\Phi^2$ represents the resistance of the structure to rotations from ideal crosslinking geometry,
 \begin{equation}
 \label{eq: K}
K_{n_b} = \frac{ n_b U}{ 1+ \frac{U}{2C}(1-\frac{n_b}{28})^2}.
\end{equation}
This elastic response of the crosslinking array to coherent rotation is straightforward to understand in the small and large $U$ limits.  The rotation of filaments from the ideal bonding state requires either the bond orientation -- as parameterized by $\delta \phi_{i, \ell}$ for crosslinked monomers --
or the torsional state of the filament within the 28-monomer repeat to adjust.  When $U/C \ll 1$ and linkers are more flexible than the filaments, this torsional load is carried by the flexibility of the $n_b$ crosslinks themselves, hence, $K_{n_b} \propto n_b U$.  For very rigid linkers, $U/C \gg 1$, bound monomers are pinned to the lattice directions the rotation of the filament section is accomplished instead by a twist distortion of the monomer segments neighboring bonds, so that $K_{n_b} \propto n_b C$.   

\begin{figure}
\centering
\center \epsfig{file= 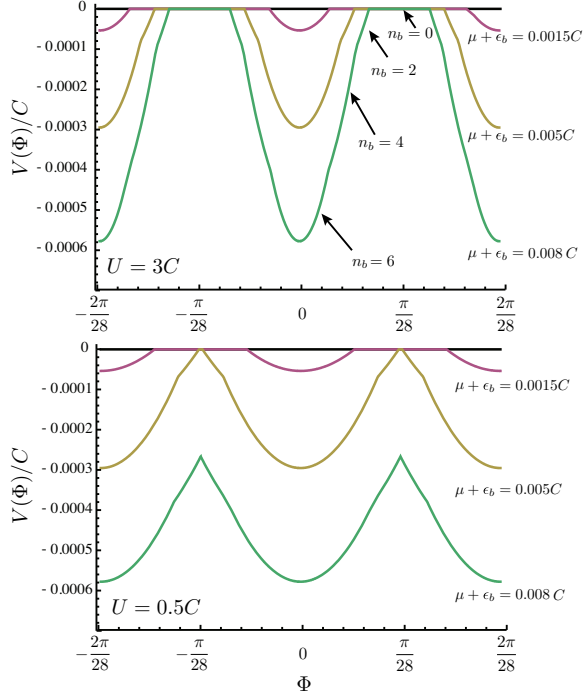, width=3.0in}
\caption{The free energy gain per monomer of crosslinking as a function of $\Phi$, the angle of coherent rotation of filaments in the bundle for different values of the linker chemical potential, eq. (\ref{eq: V}).  Highlighted in the for the rigid linker case, $U=3C$, are regions of the rotational potential where a 28-monomer repeat maintains 0, 2, 4 and 6 bonds.  The maximal number of bonds occurs for commensurate geometries with $\Phi = 2 \pi m /28$.}
\label{fig: potential}
\end{figure}

Shown for flexible and stiff crosslinkers in Fig. \ref{fig: potential}, $V(\Phi)$ functions as the ``rotational potential" describing the free energy preference for the filament structure to lock-in to a set of preferred torsional configurations due the favorable number and arrangement of bonds in overtwisted bundles.  By definition the coherent rotation varies slowly on the scale of monomers.  We take the continuum limit of our model, $\Phi(\ell)=\Phi_\ell$ and write the final form of the free energy of a bundle with $n_f$ filaments as,
\begin{equation}
\label{eq: fk}
F_{bundle}[\Phi(\ell)]= n_f \int_0^L d\ell~\Big[ \frac{C}{2} \Big(\frac{d \Phi}{d \ell}-\delta \omega_0\Big)^2 + V\big( \Phi(\ell) \big) \Big] .
\end{equation}
Written as such, the coarse-grained free energy highlights the essential frustration of parallel actin bundles.  The first term in the integrand is minimized when $d \Phi/ d z = \delta \omega_0$ and the filaments in the bundle revert to the native, 13-fold helical geometry.  Competing with this is the rotational potential, which is minimized by a constant value $\Phi = 2 \pi m /28$.  The relative importance of these competing effects is sensitive to $\mu$, which largely dictates the depth of $V(\Phi)$, but also $U$ and $C$ which together determine the relative stiffness of filaments and the pinning of rotational potential.

The effective model for parallel actin bundles, eq. (\ref{eq: fk}), is known in condensed matter contexts as the Frenkel-Kontorowa model, employed in the study of incommensurate, one-dimensional solids~\cite{bak}.  The structure and thermodynamics of the minimal energy ground states show a complex dependence on the degree of incommensurability, $\delta \omega_0$, and relative strength of the potential pinning the solid in the commensurate state (here the overtwisted -28/13 bundle) to the elastic energy of the incommensurate state (the native -13/6 symmetry).  The free-energy minimizing solutions are described the following differential equation~\cite{chaikin},
\begin{equation}
\label{eq: eom}
\frac{C}{2}\left(\frac{d \Phi}{d\ell}\right)^2 = V( \Phi)-V(0)  +\epsilon ,
\end{equation}
where $\epsilon$ is a non-negative parameter that specifies the entire rotational structure, $\Phi(\ell)$, along the bundle.  Minimizing the bundle free energy (\ref{eq: fk}) of this class of solutions yields an equation for $\epsilon$ which corresponds to the mean structure of the bundle,
\be\label{min_solution}
\delta \omega_0 = \frac{28}{\sqrt{2 C} \pi }\int_0^{ 2\pi/28} d\Phi  \sqrt{ V(\Phi)- V(0) +\epsilon} \ .
\ee
Even in the absence of thermal fluctuations, this model has a complex dependence on the binding free energy.  For sufficiently strong pinning potentials, the lowest energy solution becomes $\Phi = 0$ and $\epsilon = 0$, indicating that the bundle has locked-into the {\it commensurate phase}, here -28/13 overtwisted structure.  Below a critical depth of the pinning potential, solutions with $\epsilon > 0 $ exist indicating that $\Phi(\ell)$ has an inhomogeneous solution, which gradually unwinds to the native state.  This corresponds to the {\it incommensurate phase} of the Frenkel-Kontorowa model.  As a measure of the average rate of rotation of filament structure, we define the length ${\cal L}$ as the length along which the filament geometry unwinds by $2 \pi/28$, from one minimum of $V(\Phi)$ to the next.  From eq. (\ref{eq: eom}) this length, measured in monomer number, is computed from the integral,
\begin{equation}
\label{eq: L}
{\cal L} = \sqrt{\frac{C}{2}} \int_0^{2 \pi/28}  \frac{d\Phi}{\sqrt{V(\Phi)-V(0)+\epsilon}}.
\end{equation}
This length is related to the mean rate of filament twist by,
\begin{equation}
\langle \Delta \phi \rangle = \Omega_0 - \frac{ 2 \pi}{28} {\cal L}^{-1}.
\end{equation}
In the following section, we analyze the behavior of this order parameter, as well as the detailed structure of parallel bundles in terms of the inhomogeneous solutions for filament rotation, $\Phi(\ell)$.  
 
\begin{figure}[t]
\centering
\includegraphics[angle=0.0, width=2.6in]{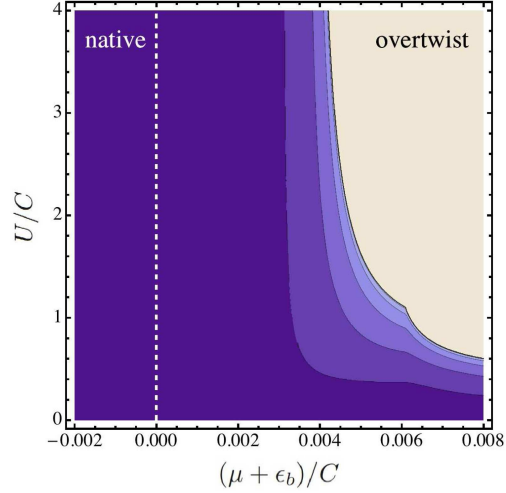} \hspace{3mm}
\caption{The twist-state phase diagram -- contours of $\langle \Delta \phi \rangle$ -- in the parameter space of $(\mu+\epsilon_b)/C$ and $U/C$.  The evolution of the stable twist state from -28/13 (overtwist) to -13/6 (native) state is observed. } 
\label{fig: phasediagram}
\end{figure}

\begin{figure}[bh]
\centering
\includegraphics[angle=0.0, scale=0.5]{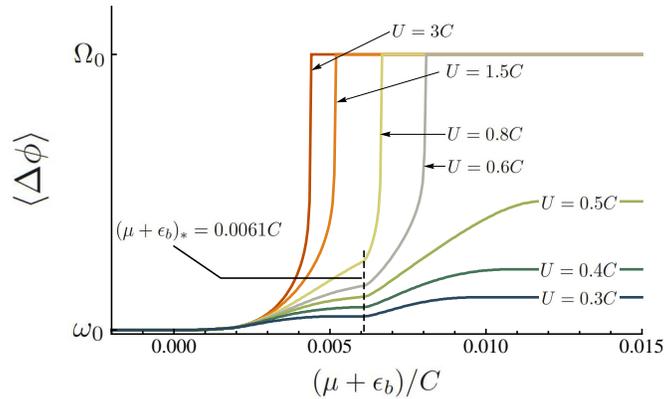}
\caption{The mean rate of twist $\langle \Delta \phi \rangle$ per monomer, from $\omega_0$ to $\Omega_0$ , is plotted as a function of the chemical potential $(\mu+\epsilon_b)/C$ for the various values of $U/C$.  The dotted line is for critical value of $(\mu+\epsilon_b)=0$, where the minimum in the binding free energy potential abruptly changes from $V(0)=0$ to $V(0) < 0$.} 
\label{fig: twist_mu}
\end{figure}

\section{Overtwist Transition}

The overtwist thermodynamics of crosslinked parallel-actin bundles predicted from the coarse-grained model is shown in Fig. \ref{fig: phasediagram}, which shows the mean filament twist in terms of crosslinker chemical potential, $\mu$, and the stiffness of crosslinking bonds, $U$.  For $\mu \leq -\epsilon_b$, crosslinks are not favored in the bundle, and hence $V(\Phi) =0$, indicating no thermodynamic preference for twist, hence, in this region $\langle \Delta \phi \rangle = \omega_0$, the native state of  actin geometry.  For all values of  $\mu > -\epsilon_b$, crosslinking is favorable and the incommensurate geometry of the -28/13 ideal crosslinker favors an overtwisted actin geometry with $\langle \Delta \phi \rangle > \omega_0$.  

The sensitivity of filament twist to $\mu+\epsilon_b$ is dramatically different for {\it stiff} and {\it flexible} crosslinking bonds.  Shown in Fig. \ref{fig: twist_mu} is the $\mu$-dependence of filament twist plotted for linker stiffnesses ranging from more rigid ($U>C$) to more flexible ($U<C$) than effective torsional resistance of 2 successive monomers.  For rigid linkers, we predict that filament twist initially increases continuously with increased $\mu$ near the onset of crosslinker binding $\mu \gtrsim - \epsilon_b$,  but upon approaching a critical value of the chemical potential, $\mu_{{\rm CI}}$, very rapidly overtwists to the -28/13 structure.  For $\mu > \mu_{{\rm CI}}$, the filament geometry locks into $\langle \Delta \phi \rangle = \Omega_0$ and exhibits no further sensitivity to the availability of crosslinking proteins.  

This singular dependence of $\langle \Delta \phi \rangle$ on $\mu+\epsilon_b$ is the signature of commensurate-incommensurate (CI) phase transition of eq. (\ref{eq: fk}), marked by a divergence of the distance over which $\Phi$ rotates by $2 \pi/28$, ${\cal L} \sim -\ln(\mu_{\rm CI} - \mu)$.  The extreme sensitivity of bundle structure to crosslinker chemical potential near the CI phase transition derives from the highly cooperative change of symmetry of low energy state of rigidly crosslinked bundles mediated by the elasticity of the filaments and the array of crosslinking bonds in the bundles.  The value of $\mu_{{\rm CI}}$ is determined by the $\mu$ at which $\epsilon \to 0$ from eq. (\ref{min_solution}), is shown in Fig. \ref{fig: phasediagram}.  In the limit of infinite linker rigidity, $\mu_{{\rm CI}} (U\gg C) \to -\epsilon_b + 0.0038C$, and this critical value shifts to larger $\mu$ as $U$ is decreased.  In contrast to the stiff linker limit, below a critical value of linker stiffness, $U_* = 0.512 C$, the overtwist shows no CI transition, and $\langle \Delta \phi \rangle$ shows a considerably reduced sensitivity to $\mu$.

What accounts for the distinction between bundling by rigid and flexible crosslinkers?  In our coarse-grained model, these differences in thermodynamic behavior ultimately derive from distinct features rotational free energy potential, $V(\Phi)$, describing the sensitivity to the free energy gained by crosslinking to rotational state of actin filaments in the bundle.  The thermodynamic preference to lock into the -28/23 structure can be crudely understood in terms of relative cost overtwisting of the filament to a constant $\Phi=0$ state, $C (\delta \omega_0)^2/2$ per monomer, and free energy cost of rotating the ideal filament structure from the commensurate state, roughly corresponding to the depth of the rotational potential, $\Delta V=V(\pi/28) -V(0)$, describing the thermodynamic cost of the non-ideal crosslinking geometry.  From eqs. (\ref{eq: V}) and (\ref{eq: K}), the narrowness of the minima of $V(\Phi)$ is determined by $K_{n_b}$, which increases monotonically with increased linker stiffness.  As shown in Fig. \ref{fig: potential}, rigid linkers only allow favorable crosslinking for a narrow range of rotations from the commensurate geometry, and hence, rotating away from $\Phi$ between minima necessarily forces filament to release its crosslinks, $\Delta V \approx (3/28)( \mu+\epsilon_b - \delta \mu_6 )$.  Thus, for sufficiently large $\mu$ this penalty outweighs the cost of overtwist, and rigid crosslinkers always lead to bundles locked-into the commensurate state.  For flexible linkers, $V(\Phi)$ is a shallow function of rotation angle, and all rotation angles up to $\Phi=\pi/28$ allow filament segments to maintain 6 crosslinks, albeit somewhat stretched from the ideal geometry.  In this case, $\Delta V \approx (3/28) U (\pi/28)^2$, which is determined entirely by the elastic cost of stretching linkers as no crosslinks need unbind to overcome the free energy barrier.  Thus, sufficiently flexible linkers allow the bundles to accommodate a maximal number of favorable crosslinking proteins even in the absence of a dramatically overtwisted structure.  Therefore, we see that $\langle n_b \rangle$ the mean number of crosslinkers per 28 monomer repeat (see Appendix \ref{app: nb}), shown in Fig. \ref{fig: bonddensity} is predicted to increase to 6 crosslinks for large $\mu+\epsilon_b$, independent of crosslink flexibility.  

\begin{figure}[t]
\centering
\includegraphics[angle=0.0,  width=2.6in]{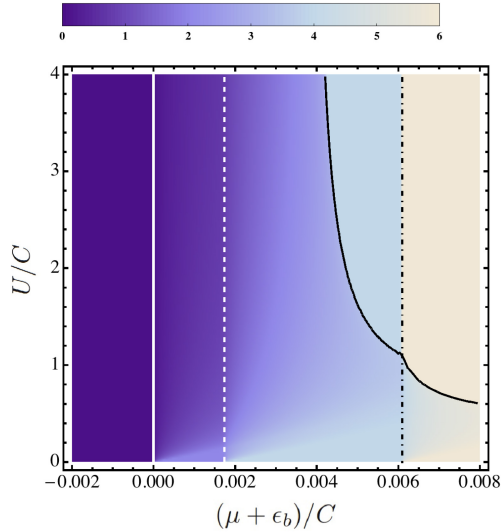}
\caption{ The mean number of crosslinkers, $\langle n_b \rangle$, per 28-monomer repeat unit.  The solid curve indicates the commensurate-incommensurate transition, while the vertical lines at   $\mu +\epsilon_b = 0, 0.0017C$ and $0.0061C$, denote abrupt transitions in minimal-energy number of crosslinkers as described by eq. (\ref{eq: V}). } 
\label{fig: bonddensity}
\end{figure}

\section{States of Intermediate Twist}

\begin{figure}
\centering
\center \epsfig{file= 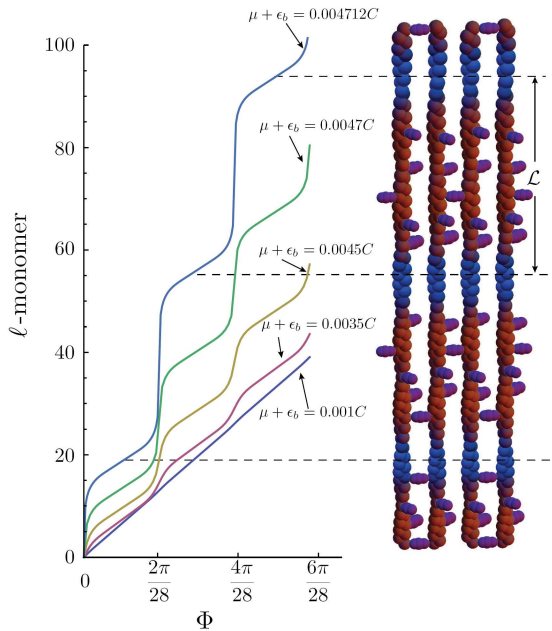, width=3.in}
\caption{ The profile of coherent rotation $\Phi$ with respect to the $\ell$ monomer is plotted at the various values of $\mu+\epsilon_b$. On the right, the inhomogeneous structure of parallel bundles, composed by alternative domains of sparsely-bound native (blue) and densely-bound, overtwist (red) symmetry, is shown schematically. }
\label{fig: profile}
\end{figure}

Upon crosslinking, a parallel actin bundle undergoes a complex structural change, from a state of filaments possessing a native helical geometry, to a fully-bundled state with an ultimate geometry that is sensitive to the flexibility of the crosslinking bonds themselves.  The one-dimensional, coarse-grained model of eq. (\ref{eq: fk}) predicts that a bundle progresses between these limiting structures via a rich pathway of intermediate states of inhomogeneous filament twist.  Shown in Fig. \ref{fig: profile} are minimal energy solutions of eqs. (\ref{eq: eom}) and (\ref{min_solution}) for the lengthscale, coherent rotations of bundles filaments, $\Phi(\ell)$, for $U=2C$.  

Beginning in the commensurate state in excess of available crosslinkers ($\mu > \mu_{{\rm CI}}$), the bundle filaments lock into the ground state of binding free energy density, $\Phi(\ell)=0$.  For $\mu$ just below this critical value, a weaker rotational potential does not hold the filament in the commensurate overtwist state along the entire filament length.  Instead, filaments untwist by rotating from one commensurate orientation to the next, say from $2 \pi m /28$ to $2 \pi (m+1)/28$, by way of rapid ``jumps" in $\Phi(\ell)$.  In the language of incommensurate solids, these anglar jumps are known as solitons or discommensurations.  In this model a discommensuration spans the cross section of the bundle, representing a domain of nearly native filament geometry,  $d \phi/ d \ell \approx \delta \omega_0$.    Over a relatively short span of roughly $\sim 15 -20$ monomers $\Phi$ rotates between two nearby minima in the binding free energy.  For rigidly crosslinked bundles, the mean number of crosslinks in these domains is significantly reduced from $n_b=6$ per 28 monomer sections, as crosslinks are predicted to unbind for at the free energy maximum of $V(\Phi)$ (see Fig. \ref{fig: potential}).  

The inhomogeneous twist and crosslinking structure of parallel bundles is shown skematically in Fig. \ref{fig: profile}.   The minimal free energy bundle configurations are described by periodic solutions for $\Phi(\ell)$, in which discommensurations have an equilibrium spacing, ${\cal L}$, along the bundle.  As this length becomes very large close to the $\mu_{{\rm CI}}$, the regions between discommensurations represent sections of overtwisted, maximally crosslinked bundles.  Hence, a surprising prediction of this coarse-grained model is that states of intermediate twist are constructed of alternating domains of sparsely-bound, native filament geometry and strongly-bound, overtwisted filament geometry.  The net rate of filament twist, $\langle \Delta \phi \rangle$, is predicted to increase as the relative proportion of overtwist to native twist filament geometry increases, as further portions of bundle are converted to highly-crosslinked, -28/13 geometry.  As shown in Fig. \ref{fig: profile}, when $\mu$ is reduced well below $\mu_{{\rm CI}}$, ${\cal L}$ decreases until successive discommensurations merge, relaxing the filament to its homogeneous state of native geometry.

\section{Discussion}

When actin filaments are bundled in parallel arrays by action of compact, globular bundling proteins, the process of crosslinking affects complex structural change of the filaments themselves, an overtwisting from a native -13/6 helix to -28/13 symmetry.  This structural change is required to maximize the number of co-oriented monomers on neighboring filaments, themselves arrayed on a hexagonal lattice.  The necessity to overtwist actin filaments upon binding of crosslinking proteins leads to a complex thermodynamic dependence of bundle properties -- mean twist and bound crosslinker density -- on the availability of crosslinkers.  As it is the bonds themselves that mechanically distort the helical structure of bundled filaments, we find that the thermodynamics of the bundling process is extremely sensitive to the rigidity of the crosslinking bonds.  A mapping of a coarse-grained free energy of bundles onto the Frenkel-Kontorowa model of incommensurate solids predicts that rigid crosslinkers ($U > U_*$) rapidly overtwist filaments via a second-order phase transition, that takes place over a very narrow range of crosslinker chemical potential.  In contrast, flexible crosslinks are predicted to be much less efficient in providing the necessary overtwist of filaments, leading to a bundling transition that takes place over a much broader range of chemical potential, and consequently, over broad range of free crosslinker concentrations.  

This distinction in crosslinker behavior between {\it rigid} and {\it flexible} crosslinkers predicted here is entirely consistent with the observed differences between crosslinking behavior between the globular crosslinking proteins, espin~\cite{purdy, shin} and fascin~\cite{claessens_pnas_08, shin}.  Based on our present study, we attribute the respectively abrupt and continuous dependence of filament overtwist observed for espin- and fascin-mediated bundles, not to differences in linker affinity, but to differences in flexibility of the crosslinking bond.  This suggests that, in part, different cells form bundles by way of different bundling proteins, in order to regulate not just the structure of bundles, but also self-assembly pathway of bundles and consequent sensitivity of these bundles to modulations in the availability of bundling proteins.  By altering the flexibility of the crosslinking agents alone, the formation of bundles can be tuned from a highly-cooperative, switch-like dependence to non-cooperative, continually-varying dependence on crosslinker concentration.  

A critical value of crosslink stiffness, $U_* = 0.512C$, separates the stiff linker behavior, which exhibits a CI transition, from the flexible linker behavior, exhibiting no overtwist phase transition.  The twist modulus of actin is measured to be $C\simeq 2.8 \times 10^{-17} ~ {\rm J} \simeq 6900 ~ k_B T$ (units of angular distortion per successive monomer pair)~\cite{tsuda}.  If we attribute the elastic cost of distorting binding directions to a change of crosslinker length, we may estimate an effective spring constant for linkers with this critical stiffness $k_* \approx 500 ~ {\rm pN/nm}$, based $a \simeq 3.75~{\rm nm}$, half the diameter of actin~\cite{holmes}.  This is considerably more rigid than the bonds provided by the relatively extended crosslinkers in acto-myosin bundles~\cite{howard}, or the much larger bundling proteins, filamin and $\alpha$-actin~\cite{ferrer}, all of order $0.1-10 ~{\rm pN/nm}$.  The stiffness of crosslinking bonds provided by small, globular bundling proteins is not a well characterized quantity, though it seems reasonable that $U \sim C$, since the torsional flexibilty of the actin filaments requires distortion of specific protein-protein bonds between actin monomers that may not be wholly unlike the distortion of the bonds between actin monomers and bundling proteins of roughly the same size.   Based on a mechanical model of crosslinker shear~\cite{heussinger}, measurement of the effective bending stiffness of parallel bundles have been used to infer a shear stiffness of fascin crosslinks~\cite{claessens_nat_06} that is 3-4 orders of magnitude below this estimate for $k_*$, suggesting that fascin should fall well within the flexible linker regime.  Notwithstanding the poor understanding of the mechanics of crosslinking at the protein scale, the classification of fascin as a {\it flexible} bundling protein is consistent with its role as a primary crosslinker in filapodial bundles~\cite{faix}.  Fascin has been observed to transiently bind and unbind along filapodia {\it in vivo}, diffusing at a remarkable rate within bundles~\cite{aratyn}.  In the context of our present study, we note that flexible linkers bind and unbind in an essentially independent, non-cooperative manner.  In contrast, we expect the kinetics of rigid crosslinkers to be dramatically slower than flexible linkers due to the cooperative organization of many linkers and coherent restructuring of filaments required for each additional rigid crosslinking bond.    

We conclude by discussing the role of two effects not included in our model crosslinked parallel actin bundles:  thermal fluctuations of monomers and crosslinks and the global twisting of {\it finite} bundles.  In the present model, the complex thermodynamic binding properties of crosslinking proteins derives purely from an elastic frustration between optimal linker and filament geometries, neglecting thermal fluctuations of monomers and bonds.  In ref. \cite{shin} we studied this lattice model of parallel bundles in the presence of strong thermal fluctuations of filament orientation and crosslinker position.  In this regime, the thermodynamic distinction between by rigid and flexible bundling transition is maintained: rigid crosslinkers affect a phase transition between native and overtwist, while the flexible linkers continuously overtwist filaments upon increasing crosslinker binding.  Given the high torsional modulus~\cite{tsuda}, one expects a torsional persistence length of free, unbundled filaments, $\xi_t = 2 C/k_B T \simeq 13,000 ~{\rm monomers}$ defined by $\langle \cos (\phi_{\ell_0} - \phi_{\delta \ell+\ell_0}) \rangle = \cos( \omega_0 \delta \ell) e^{-|\delta \ell |/\xi_t}$.  On the scale of the coarse-grained repeat lengths of our model, 28 monomers, thermal fluctuations of filament twist are therefore extremely modest.  However, we do expect that the entropy associated with distributing crosslinkers at different positions in bundles to quantitatively modify the predictions of our present study, although modestly.  In particular, the non-analytic dependence of the binding free energy, $V(\Phi)$, on chemical potential and filament rotation is smoothed out by the equilibrium distribution of crosslinks among the many competing crosslinking geometries (see supporting information).  Therefore, certain sharp features of binding thermodynamics predicted by this ``zero temperature" theory are somewhat smoothed out when considering these fluctuations.  In particular, the number of bonds in the minimal free energy states changes abruptly from $n_b =0$ to 2 at $\mu +\epsilon_b = 0$,   $2$ to 4 at $\mu +\epsilon_b = 0.0017C$ and 4 to 6 at  $\mu +\epsilon_b = 0.0061C$, leading to kinks in the predicted equations of state for $\langle n_b \rangle$ and $\langle \Delta \phi \rangle$ vs. $\mu$ (Figs. \ref{fig: twist_mu} and \ref{fig: bonddensity}), which only appear in this limit.  Nonetheless, we find that presence of these fluctuations does not eliminate the sharp CI native/overtwist transition that occurs for linkers above a critical stiffness, $U_* = 0.512C$, where $\partial \langle \phi \rangle /\partial \mu$ diverges.  That is, the sharp/smooth overtwist transition for rigid/flexible crosslinkers is robust feature of the geometrical frustration in parallel actin bundles, insensitive to the presence of thermal fluctuations of bundles.  

In our model, we considered parallel actin bundles of essentially unlimited length and width, focussing on the internal reorganization of filaments within the bundle.  Cells maintain a careful control over the size of bundles, most notably in the mechanosensory hair bundles of the cochlea~\cite{tilney_saunders}.  Understanding the physical mechanisms that underlie the {\it in vivo} control of bundles on multi-filament lengthscales ($\sim 0.1 -10 ~{\rm \mu m}$) remains an outstanding puzzle.  {\it In vitro} studies of bundle formation observe that the lateral diameter is indeed sensitive to the concentration of crosslinkers in fascin/actin solutions~\cite{haviv}.  Claessens {\it et al.} suggest that the observed coincident overtwisting of filaments and growth of bundle diameter upon increased fascin binding implies that overtwist plays a role in limiting the the lateral assembly of actin bundles~\cite{claessens_pnas_08}. Indeed several theoretical studies~\cite{turner, grason_prl_07, grason_pre_09, hagan} demonstrate that a tendency of filaments to globally twist around the central axis of the bundle leads to thermodynamic frustration that ultimately limits the equilibrium diameter of bundles.  This mechanism is believed to play a role in the self-limited size of fibrin bundles~\cite{weisel}, which are clearly observed to globally twist in electron microscopy studies.   A similar mechanism may be at work in fascin-mediated bundles, as the intra-filament and inter-filament twist of helical filaments are geometrically coupled~\cite{neukirch}.  To understand the relationship between crosslinker binding and bundle size, it is therefore necessary to consider a more complex model of protein-mediate parallel bundle formation, in which filaments trade the elastic cost of filament overtwist for the elastic cost of globally twisting the entire bundle.   Though it is plausible that the tendency to overtwist individual filaments also implies a tendancy to twist the entire filament lattice in a bundle, it remains to open to question whether the elasticity of crosslinking bonds is of sufficient rigidity to mechanically bend filaments into the complex superhelical structures that frustrate and limit self-assembly.

\acknowledgements{The authors are grateful to G. Wong for many essential insights and stimulating discussions.  This work was supported by the NSF under awards NSF DMR-0820506 and Career DMR 09-55760.  The authors acknowledge the hospitality of the Aspen Center of Physics, where some of this work was completed.

\appendix
\section{Rotational potential}
\label{app: V}

We drive the explicit form of the potential $V(\Phi)$ in eq~(\ref{eq: V}) by considering a 28-monomer repeat length of filament in which $n_b$ of the ideal bond structure form with neighboring filaments.  We describe the angles of this initial state with $n_b$ monomers aligned to six-fold lattice directions by the angles $\phi^{n_b}_\ell$, and each of these states has a composite -28/13 symmetry.  As bonds are elastic and commensurate orientation varies along the bundle length, the perfect orientation of these bonds will relax to a low energy configuration.  

To determine the relaxation of orientation within a 28-mononer unit cell, we introduce two angular deviations: $\delta \phi_{b}$, the crosslinker deformation from the perfectly aligned state at the crosslinked layers and  $\delta \phi_f$ the adjustment of unbound, or free, monomers.    At layers bound by crosslinks, say $\ell_0$, the free energy of bonding per monomer from eq. (\ref{eq: binding}) is simply $-(\mu+ \epsilon_b)/2+U(\delta \phi_b)^2/2$, since each crosslink is shared by 2 filaments.  Additionally, the twist elastic distortion between the $\ell_0$ and $\ell_0 \pm1$ monomers from eq. (\ref{eq: twist}) contributes, $C(\Delta \tilde{\phi}^{n_b}_{\ell_0} + \delta \phi_b - \delta \phi_f)^2/2+C(\Delta \tilde{\phi}^{n_b}_{\ell_0+1} - \delta \phi_b + \delta \phi_f)^2/2$ to the free energy of binding, where $\tilde{\phi}^{n_b}_\ell = \phi^{n_b}_\ell - \Omega_0 \ell$.  The twist elastic energy of the remaining unbound monomer sections is simply, $C(\Delta \tilde{\phi}^{n_b}_{\ell})^2/2$.  Summing these contributions, we have the free energy of binding within a 
28-monomer section per filament in the $n_b$ bond state,
\be\label{eq: F_nb}
F^{n_b}_{28} (\delta \phi_b, \delta \phi_f) = \frac{n_b}{2}\left( U(\delta \phi_b)^2+2C (\delta \phi_f -\delta \phi_b)^2-(\mu +\epsilon_b)\right)+ \frac{C}{2}\sum_{ \ell=0}^{28} (\Delta \tilde \phi^{n_b}_{\ell})^2  \ .
\ee
Since $\Phi$ is the coarse-grained coherent rotation of all these deformation in a 28-monomer unit, we have the following relation,
\be
\Phi = \frac{n_b\delta \phi_b+(28-n_b)  \delta \phi_f}{28} \  .
\ee
Minimizing eq.~(\ref{eq: F_nb}) with respect to the monomer adjustment angle, $\delta \phi_b$, for fixed $\Phi$ we find the minimal energy bond angle deviation is 
\be \label{eq: phi_b}
\delta \phi_b = \frac{\Phi}{\frac{ U}{2 C} (1-\frac{n_b}{28})^2+1} \ .
\ee
Note that in the limit of rigid crosslinkers, when $U \gg C$, the bound monomers are pinned to the lattice directions, $\delta \phi_b = 0$.  The resulting minimal free energy of the $n_b$-monomer state rotated to an angle, $\Phi$, is
\be
\label{eq: Vnb}
F^{n_b}_{28} (\Phi)=\frac{K_{n_b}}{2} \Phi^2 -\frac{n_b}{2}(\mu -\mu_c(n_b))
\ee
where  $K_{n_b}$ is defined as in eq.~(\ref{eq: K}) and $\mu_c(n_b)=-\epsilon_b+C\sum_{\ell=0}^{28} (\Delta \tilde{\phi}^{n_b}_{\ell})^2 /n_b$.  The rotational potential is determined by minimizing $F^{n_b}_{28}(\Phi)$ over all possible bond configurations within the 28 monomer repeat for a given value of $\Phi$, according to eq. (\ref{eq: V}).  Though are a total of states considered within this class, for a given bond number a single bond configuration minimizes the elastic ``offset" energy, $C\sum_{\ell=0}^{28} (\Delta \tilde{\phi}^{n_b}_{\ell})^2$ (see table II  in the supporting materials).

\section{Mean crosslink number}
\label{app: nb}
At a given coherent rotation $\Phi$,  the optimal number state of crosslinkers, $n_b(\Phi)$, is defined as the number state which minimizes eq.~(\ref{eq: Vnb}) over all the states of perfectly aligned crosslinking, from $n_b=0$ to $n_b=6$.  As shown in examples of  stiff crosslinkers in Fig.~\ref{fig: potential},  the minimized potential $V(\Phi)$ possesses the subset of certain number states over the periodic range of $\Phi$.  

In the minimal free energy configurations, $\Phi(\ell)$ rotates through multiple angles along the bundle corresponding to the solution of eq. (\ref{eq: eom}),
\be
\ell = \int_0^{\Phi(\ell)} d \Phi' \Big(\frac{d \Phi'}{d z}\Big)^{-1} .
\ee
We calculate the mean number of crosslinkers in a 28 monomer repeat $\langle n_b \rangle$ as following,
\be
\langle n_b \rangle = \frac{1}{\mathcal L}\int_0^{2\pi/28}d\Phi  \frac{n_b(\Phi) }{\sqrt {2 ( V (\Phi)-V(0) +\epsilon)/C}}  \ .
\ee

\newpage

\noindent {\bf \large Supporting Material}

\vspace{1cm}

\noindent{\bf S1. Ideal crosslinking geometry of parallel actin bundles} 
\vspace{0.01cm}
\\

\renewcommand{\theequation}{S\arabic{equation}}
\addtocounter{equation}{-2}
Here, we describe the details of the optimal packing calculation of our overtwist groundstate as shown in Fig. 2.
The ideal crosslinking configurations should fulfill the following conditions: 1) maximize crosslinking bonds that are perfectly aligned to the hexagonal packing directions and 2) minimize the twist distortion from the native $-13/6$ symmetry.  To search for the optimal crosslinking configuration of parallel actin bundles, we first consider a single filament configuration with the commensurate helical geometry of $-6m/n$ symmetry, in which each successive monomer is rotated by $-2\pi n/(6m)$, where both $n$ and $m$ are integers.  This class of helical symmetry allows a subset of monomers align up with one of the sixfold lattice directions.   In the block of $m$ consecutive monomers with $-6m/n$ symmetry, two monomers at the boundary of the block make perfect alignments with the packing direction:  for example, if the first monomer starts at $\phi=0$, the $m$th monomer lines up with  $\phi=-2\pi n/6$. Among these helical symmetries, we consider the configurations parameterized by a pair of integers, $(n,m)$, which minimize the angular deviation from the native symmetry.   In Fig.~\ref{fig: angle_diff}, we plot the angle difference from $-13/6$ symmetry for $-6m/n$ symmetry up to $m=7$.  As shown in Fig.~\ref{fig: angle_diff}, it is found that $-24/11$ and $-30/14$ -- respectively, under- and over-twisted relative to the native state -- symmetry are closest to the native state, corresponding to $m=4$ and $m=5$, respectively.  In our analysis, we focus on these two commensurate helical geometries.

We now construct composite structures of a single filament, which consists of $N_4$ numbers of 4-monomer sections with $-24/11$ symmetry and $N_5$ numbers of 5-monomer sections with $-30/14$ symmetry.   For $-24/11$ symmetry, the 4-monomer section makes $- 2 \pi (11/6)$ rotation, while for $-30/14$ symmetry, the 5-monomer section makes $-2\pi (14/6)$ rotation.   In order to make full $N$ turns, $N_4$ and $N_5$ must satisfy the following condition: $-4N_4(11/24)-5N_5(14/30) =N $ per $M$ monomers, where $M=4N_4+5N_5$ and $N$ is a negative integer for left-handed turns. As an example, the composite structures of a single filament with 28-fold symmetry are displayed in Fig.~\ref{fig: composite}, showing the longitudinal arrangements of blocks, such as 445555 and 554455 in (a) and (b), and their top views with the orientations of crosslinkers in (c) and (d), together with the crosslinker layer numbers. In table I, we present the class of composite structures, up to 102 monomers per repeat unit, that are used in this analysis.

The remaining task is to build three-dimensional bundles by tiling the hexagonal lattice with the single-filament composite motifs for a given repeat length, $M$, and search for a state that maximizes the crosslinker density (defined as the ratio between the number of co-oriented monomers and the number of monomers in a bundle). Note that crosslinking only occurs between co-oriented monomers.  For this task, we use a Monte Carlo method and find an upper bound of crosslinker density for a given symmetry. The initial configuration of the system is set by arranging filaments in the hexagonal array ($5\times5$ rows).  Each filament has one composite state of longitudinal block arrangements and one orientational state, which are randomly chosen from the block permutations of $(N_4+N_5)!/N_4!/N_5!$ states (all possible block translations and rearrangements along filament) and the six orientational states of hexagonal lattice directions.  In the MC step, we make discrete rotational trial moves of filament by $\pm 60^{\circ}$ around its axis, as well as block translational trial moves, which are accepted if the trial moves increase the number of crosslinkers (i.e., the co-oriented monomers) and otherwise, rejected. The system is let to equilibrate until the crosslinker density is saturated, which usually requires from 1000 to $5 \times10^6$ MC trial movements, depending on the number of monomers in the filament.  To calculate an upper bound on the maximum crosslinker density allowed by a given composite structure, we count the number of all co-oriented monomers belonging to the $3\times 3$ inner filaments, embedded within $5\times 5$ filament lattice.  For a given finite size of lattice, this counting provides the upper bound of crosslinker density for an arbitrary large size of lattice, because we assume that this local packing arrangement may be continued over a larger region of the bundle.  In fact, the crosslinkers along the boundary may be frustrated, resulting in an ultimately lower density of crosslinks upon considering a larger region of the bundle.  Therefore, the number of co-oriented monomers for inner filaments serves as an upper bound for the maximum crosslinkers density allowed for a given symmetry.  The MC procedure is repeated 1000 times for a given $N_4$ and $N_5$ to find the crosslinker configuration that gives the maximizes the number of co-oriented monomers.

The upper bounds for crosslinker density are displayed in the chart of Fig.~\ref{fig: mc_bonds}.  Up to 102 monomer repeat units, we find that the groundstate geometry of 28-fold symmetry provides the maximum crosslinker density, that is, it saturates its upper bound with 6 bonds for every 28 monomers on every filament.  However, for filament lattices of only $5\times 5$ rows, the upper bound on the crosslinker density of 13-fold symmetry is  very close to that of optimal packing geometry.   For the comparison with the optimal packing geometry of $-28/13$ symmetry (see Fig.~2 in the text or Fig.~\ref{fig: rotation}), we also display the crosslinking configuration for $-13/6$ symmetry in Fig.~\ref{fig: 13_6}, which has an upper bound of crosslinking density of $0.213675$.  Hence we performed more exhaustive searches on these two symmetries, by increasing the bundle size up to $10\times 10$ to obtain tighter upper bounds for these structures, as shown in Fig.~\ref{fig: 13vs28}.
As the number of rows increases, the configuration with $-28/13$ composite filaments exhibits a constant crosslinkers density of $0.214286$, while the crosslinking of the $-13/6$ becomes increasingly frustrated, and ends up with a significantly reduced upper bound on the crosslinker density, compared to that of $-28/13$ structure, for $10\times10$ filament rows, the upper bound of the density of bonds is ultimately reduced to 0.198317 (see also the dark bar in Fig.~\ref{fig: mc_bonds}). The reduced value in the larger $-13/6$ bundles can be explained by the fact that, as the lattice size increases, the number of the ``assumed" crosslinkers on boundary filaments decreases, which in turn reduces the upper bound. In the optimal crossliking geometry of $-28/13$ symmetry, unlike to the configuration with $-13/6$ symmetry, all the monomers aligned with the six-fold lattice directions are being fully consumed and involved in crosslinking.   Note that $-28/13$ structure can tile a hexagonal lattice of arbitrary size, maintaining this maximum bond density.

\vspace{1cm} 
\noindent{\bf S2. $2\pi/28$ rotational symmetry of the groundstate} 
\vspace{0.01cm}
\\

The optimal packing geometry of crosslinkning with $-28/13$ symmetry possesses the $2\pi/28$ rotational symmetry.
Here, we demonstrate that under a coherent rotation by $2\pi/28$, the bundle recovers the same crosslinking pattern of $-28/13$ overtwist groundstate, and hence, the free energy is degenerate under this rotation.  Starting with the overtwist groundstate as shown in the left panel in Fig.~\ref{fig: rotation}, we take the following steps: 1) remove all crosslinkers,  incurring a bond free-energy penalty, $+\Delta F_b$; 2) rearrange all the monomers into the homogeneous twist state of $-28/13$ symmetry by lowering twist energy, $-\Delta E_{tw}$.  Notice that at this stage, monomers at layer 0 and 14 are co-oriented along the horizontal lattice directions in Fig.~\ref{fig: rotation}.  Now, 3) rotate the entire filaments by $2\pi/28$, which brings monomers at layer 1 and 15 into perfect registry on neighbor filaments; 4) rearrange all monomers back to the original composite configurations by paying the same amount of twist energy that we gained, $\Delta E_{tw}$;  5) replace the crosslinkers, regaining the same bond energy, $-\Delta F_b$.  As a result of this coherent rotation, the layers associated with crosslinkings are changed as illustrated in the right panel in Fig.~\ref{fig: rotation}, but the original crosslinking geometry and the total energy within 28 monomer repeat unit are preserved, albeit discreted rotated.  Note that this entire procedure is identical to a one-layer translation along the bundle axis, followed by shifting the planar unit cell to the right by one-filament spacing.

\vspace{1cm}
\noindent{\bf S3. Thermal fluctuations at finite temperature} 
\vspace{0.01cm}
\\

Our model of the commensurate-incommensurate transition described in the main text considers purely the elastic energy of filaments and crosslinkers, therefore it does not include the effect thermal fluctuations of monomers and crosslinker distributions.    We here include these effects by considering all possible crosslinking geometries with a 28-monomer coarse-graining unit, and computing the probability of each state proportional to $e^{-\beta V(\Phi)}$ at finite temperature, $\beta =k_BT$.   We derive the effective pinning potential per monomer at the given temperature by considering the crosslinking statistics of each filament in bundle cross-section independently:
\bea \label{eq: V_beta}
V_{\beta}(\Phi) =-k_BT \ln \left[ 1+ z \sum_{s_1} \exp(-\beta V_{s_1}(\Phi)) +z^2 \sum_{s_2} \exp(-\beta V_{s_2}(\Phi)) \right. \nonumber\\
+ z^3 \sum_{s_3} \exp(-\beta V_{s_3}(\Phi)) +z^4 \sum_{s_4} \exp(-\beta V_{s_4}(\Phi))  \nonumber\\
\left. + z^5 \sum_{s_5} \exp(-\beta V_{s_5}(\Phi)) +z^6 \sum_{s_6} \exp(-\beta V_{s_6}(\Phi)) \right]/28 \ ,
\eea
where the fugacity is given by $z=\exp (\beta (\mu+\epsilon_b)/2)$ and the summation runs over $s_{n_b}$, all possible bond distributions for given $n_b$.  Here, the potential for each state is defined by 
\be
V_{s_{n_b}} =\left[\frac{n_b\mu_c(n_b, s_{n_b})}{2} + \frac{K_{n_b}}{2} \Phi^2 \right] \ .
\ee
Notice that the geometry of crosslinking requires us to consider a particular set of 19 correlated configurations of monomer orientations and crosslinker distributions in the 28-monomer span. In table II, we present the possible crosslinking  configurations for given $n_b$ and the corresponding the elastic offset energies, $\mu_c(n_b, s_{n_b})$.   In Fig.~\ref{fig: potential_finiteT}, we plot eq.~(\ref{eq: V_beta}) revealing only modest quantitive differences in comparison to the ``zero temperature" potential.  These curves are quantiatively similar, though the sharp features associated with the changes in the minimal free energy bond pattern have been smoothed out by thermal fluctuations between competing, nearly degenerate, bond configurations.    To calculate the twist-state phase diagram (Fig.~\ref{fig: finiteT}) including these finite temperature fluctuations, we set $\beta C=6900$ based on the known torsional stiffness of actin filaments $C \simeq 8 \times 10^{-26} \,\rm N m^2$ per the monomer spacing of 2.8 nm. 
In Fig.~\ref{fig: bonddensity_finiteT}, the mean number of crosslinkers per 28-monomer span is also calculated by introducing the bond crosslinker density operator  as follows:
\bea
n_{\beta}(\Phi) =  56 \frac{\partial V_{\beta} (\Phi) }{\partial \mu}  \ ,
\eea
along with the mean rotation profile, $\Phi(\ell)$, formula of eq. (B2).  Including the thermal fluctuations of the distribution of bonds produces a phase-diagram that is quantitatively similar to the ``zero-temperature" results presented in the text.  Note, however, that the sharp features, kinks, in the $\langle \Delta \phi \rangle$ and $\langle n_b \rangle$ dependence that derive from abrupt changes in minimal-energy number state of crosslinkers that we saw  in ``zero temerapature" calculation (see Figs.~ 4 and 6 in the text) are smoothed out.   In contrast, the sharp commensurate-incommensurate transition from the native to overtwist state is preserved at finite temperature, as the twist-suspectibility, $\partial \langle \Delta \phi \rangle / \partial \mu$, necessarily diverges as the ``lock-in" state is approached, indicating a second-order phase transition.  

\newpage

\begin{table} 
\centering
\begin{tabular}{|c||c|c| c|}
\hline
$(N_4, N_5)$ & repeat length $M$  & helical symmetry & overtwist (radians) \\
\hline \hline
$(2,1)$ & 13 &  $-13/6$ & 0.000000\\
$(6,0)$ & 24 & $-24/11$ & -0.020138\\
$(2,4)$ & 28 & $-28/13$ & 0.017261 \\
$(4,5)$ & 41 & $-41/19$  & 0.011788\\
$(10,2)$ &50 &	$-50/23$  &	-0.009666 \\
$(6,6)$ & 54 & $-54/25$ & 0.008950\\
$(14,1)$ & 61& $-61/28$ & -0.015847 \\
$(12,3) $ &63 &$-63/29$ & 	-0.007672	 \\
$(8,7)$ & 67 & $-67/31$ & 0.007214 \\
$(6,9)$ & 69 & $-69/32$ & 0.014009\\
$(14,4)$ & 76 & $-76/35$ &-0.006359 \\
$(10,8)$ & 80 & $-80/37$ & 0.006042 \\
$(20,1)$ & 85 & $-85/39$ &-0.01705 \\
$(18,3)$ &87 & $-87/40$& -0.011111 \\
$(16,5)$ & 89 & $-89/41$ &-0.005431 \\
$(12,9)$ & 93 & $-93/43$ & 0.005197 \\
$(10,11)$ & 95 & $-95/44$ & 0.010175\\
$(8,13)$ & 97 & $-97/45$ & 0.014948 \\
$(22,2)$ & 98 & $-98/45$ & -0.014796 \\
$(18,6)$ & 102 & $-102/47$ & -0.004738
\\\hline
\end{tabular}
\caption{The composite structures of a single filament constructed by $N_4$ and $N_5$ blocks with $-24/11$ and $-30/14$ symmetry, respectively.  We analyze up to 102 monomer repeat length.}
\label{tab:table}
\end{table}

\renewcommand{\thefigure}{S\arabic{figure}}
\addtocounter{figure}{-7}
\begin{figure}
\centering
\includegraphics[angle=0.0, scale=0.5]{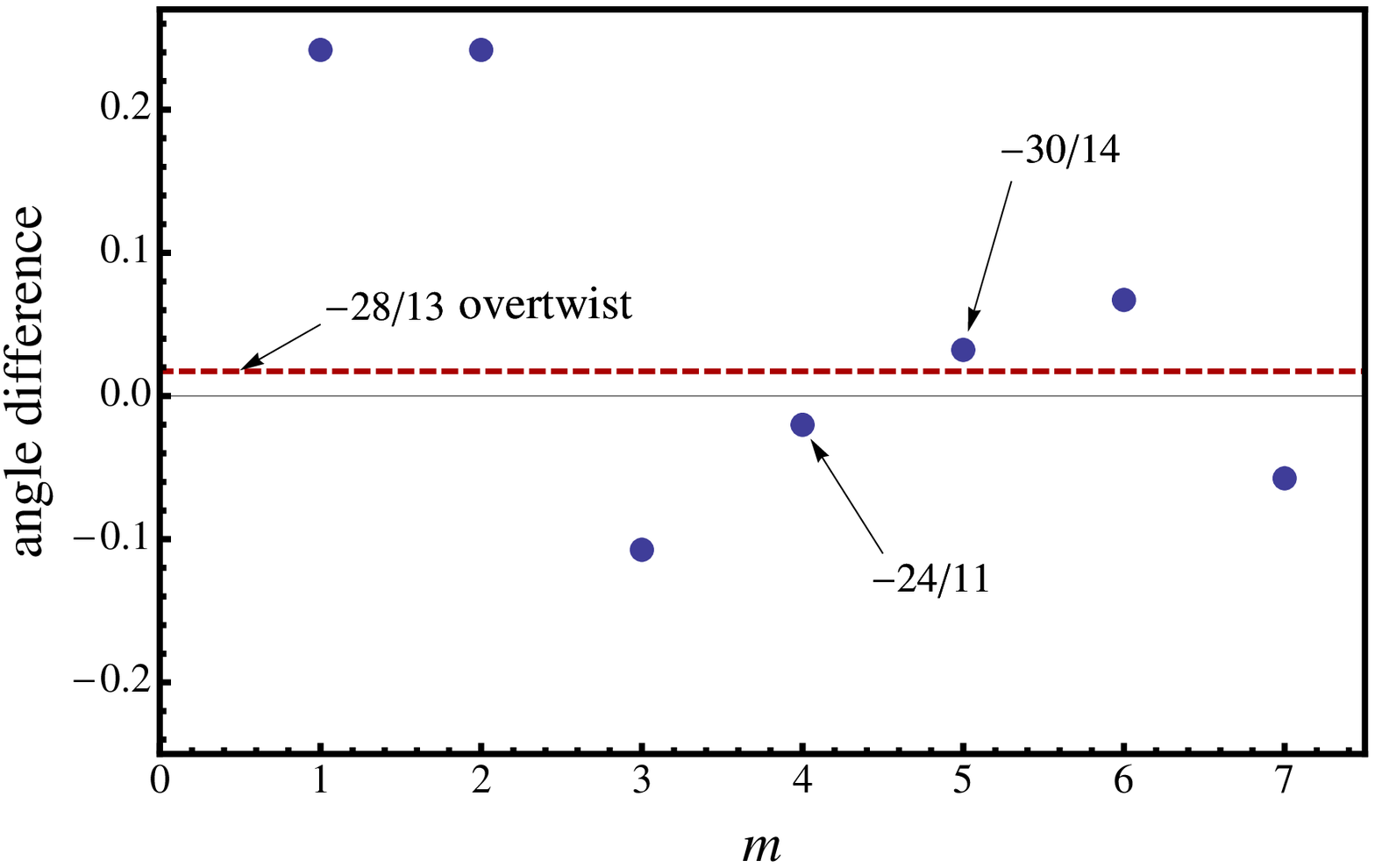}
\caption{The angle difference per monomer of $-6m/n$ symmetry from the native $-13/6$ symmetry.  $-24/11$($m=4$) and $-30/14$($m=5$) symmetry are nearest to the native state of twist. The red dashed line is the angle deviation of $-28/13$ symmetry. } 
\label{fig: angle_diff}
\end{figure}

\begin{figure}
\centering
\includegraphics[angle=0.0, scale=0.25]{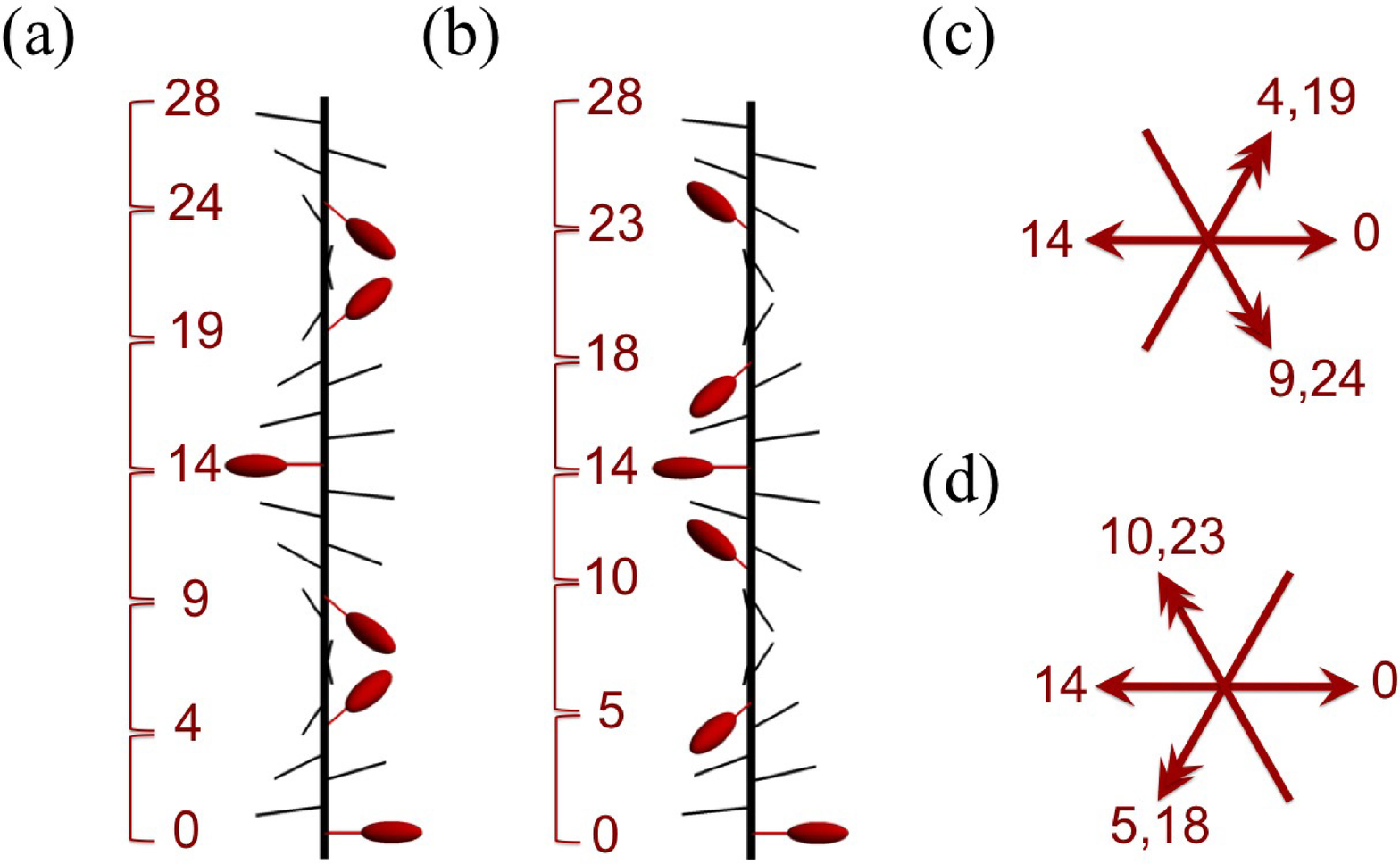}
\caption{The structural details of composite structure for 28-monomer repeat with $N_4=2$ and $N_5=4$.    The side views of filament composed by block arrangements of 445555 and 554455 are shown in (a) and (b), respectively and their top views are shown in (c) and (d).  The layer numbers are the location of crosslinkers. } 
\label{fig: composite}
\end{figure}

\begin{figure}
\centering
\includegraphics[angle=0.0, scale=0.5]{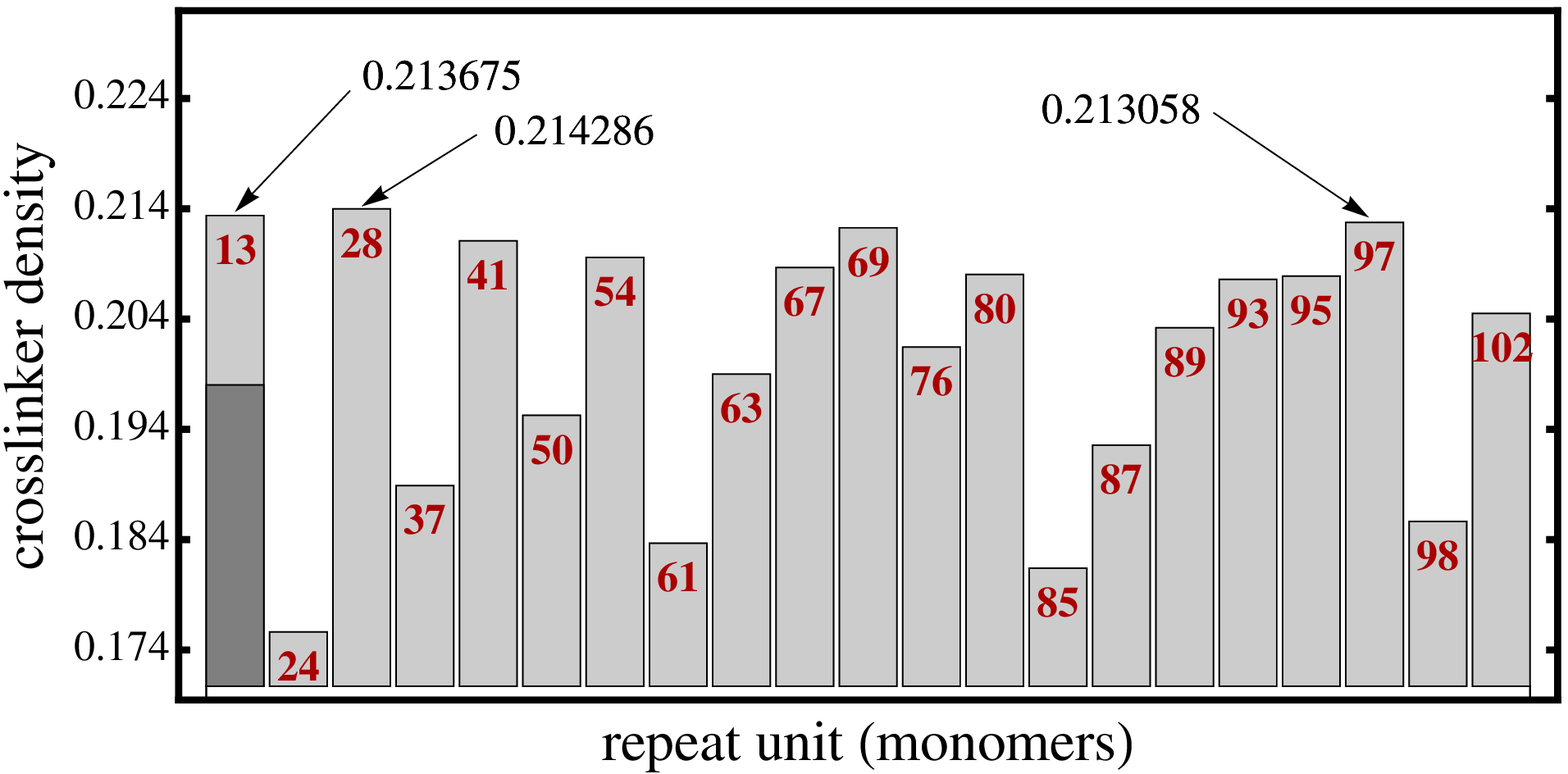}
\caption{The upper bound of crosslinker density for various symmetries up to 102 monomers are presented in bar charts.  The results are searched from a Monte Carlo calculation for $5 \times 5$ bundle size.  The 28-fold symmetry is found to be the structure with maximum crosslinker density, 6/28.  For 13-fold symmetry, the upper bound crosslinking density, $0.198317$, from $10\times 10$ row MC calculation is marked with a dark gray bar. } 
\label{fig: mc_bonds}
\end{figure}

\begin{figure}
\centering
\includegraphics[angle=0.0, scale=0.5]{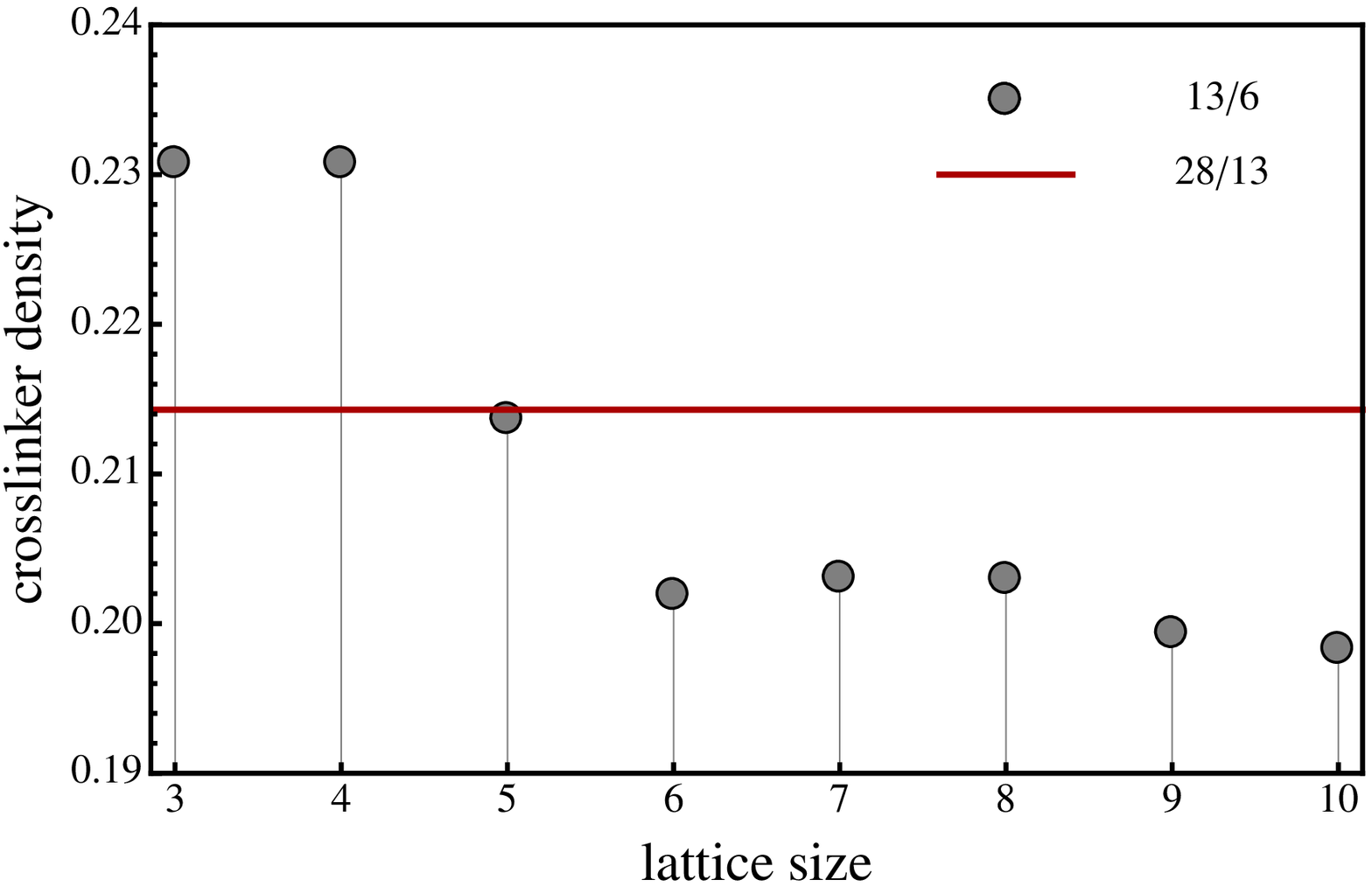}
\caption{The explicit comparison between $-13/6$ and $-28/13$ symmetry with respect to the row number in MC bond calculation.  Note that 28-fold symmetry shows a constant crosslinker density (saturating the upper bound, 6/28), while the upper bound decreases for the 13-fold symmetry as the row number grows.} 
\label{fig: 13vs28}
\end{figure}

\begin{figure}
\centering
\includegraphics[angle=0.0, scale=0.3]{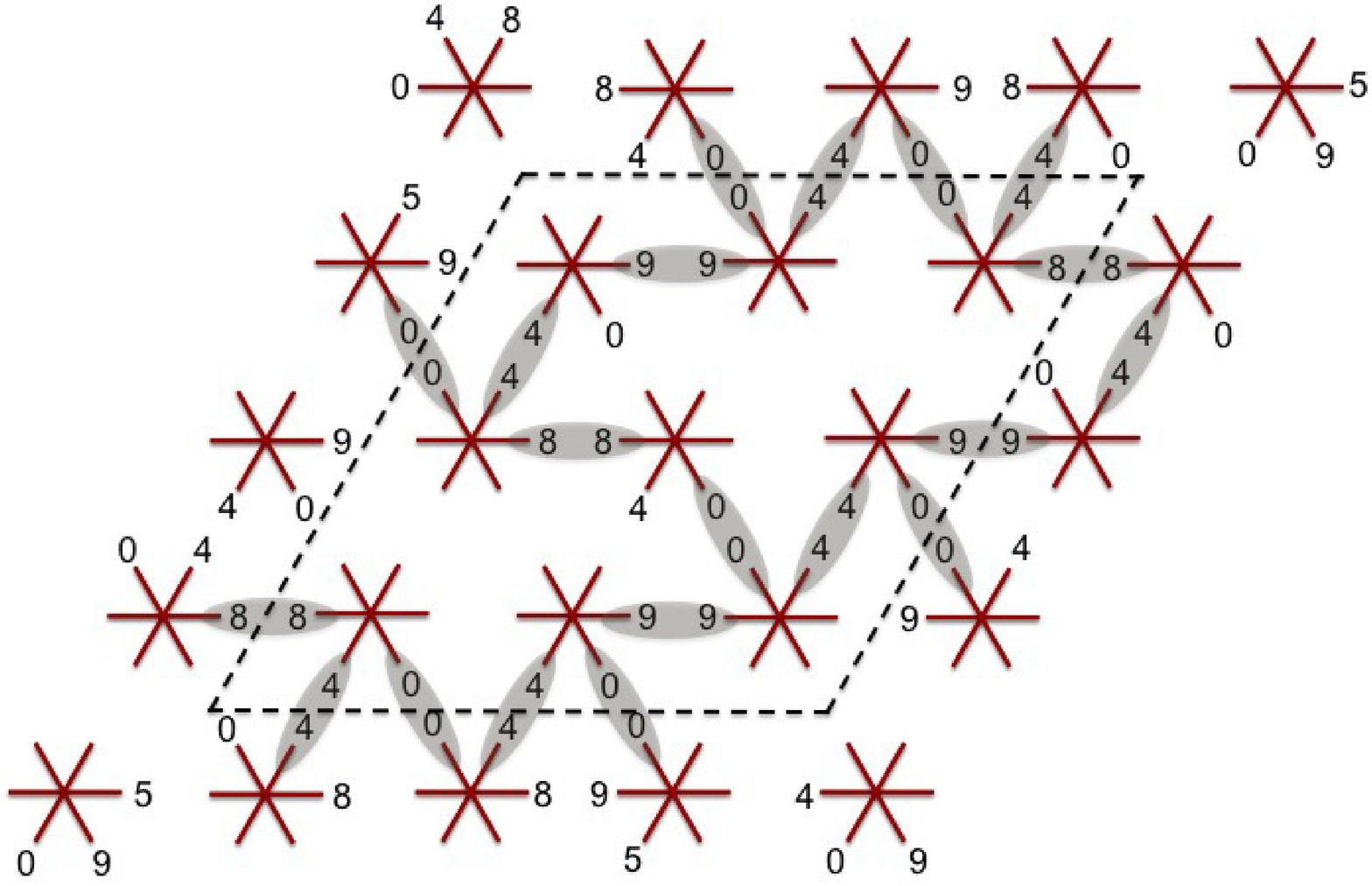} 
\caption{The maximum crosslinking configuration for $-13/6$ symmetry by Monte Carlo calculation.  We calculate an upper bound on the crosslinker density by counting crosslinks formed inside the dashed line.  Note that the crosslinkers at the boundary are share with filaments outside the cell, contributing only half a crosslink to the upper bound counter.  For a $5\times 5$ filament structure, the upper bound of crosslinker density is $(6 \times 2 +13)/ (13\times 9)=0.213675$.}  
\label{fig: 13_6}
\end{figure}

\begin{figure}
\centering
\includegraphics[angle=0.0, scale=0.25]{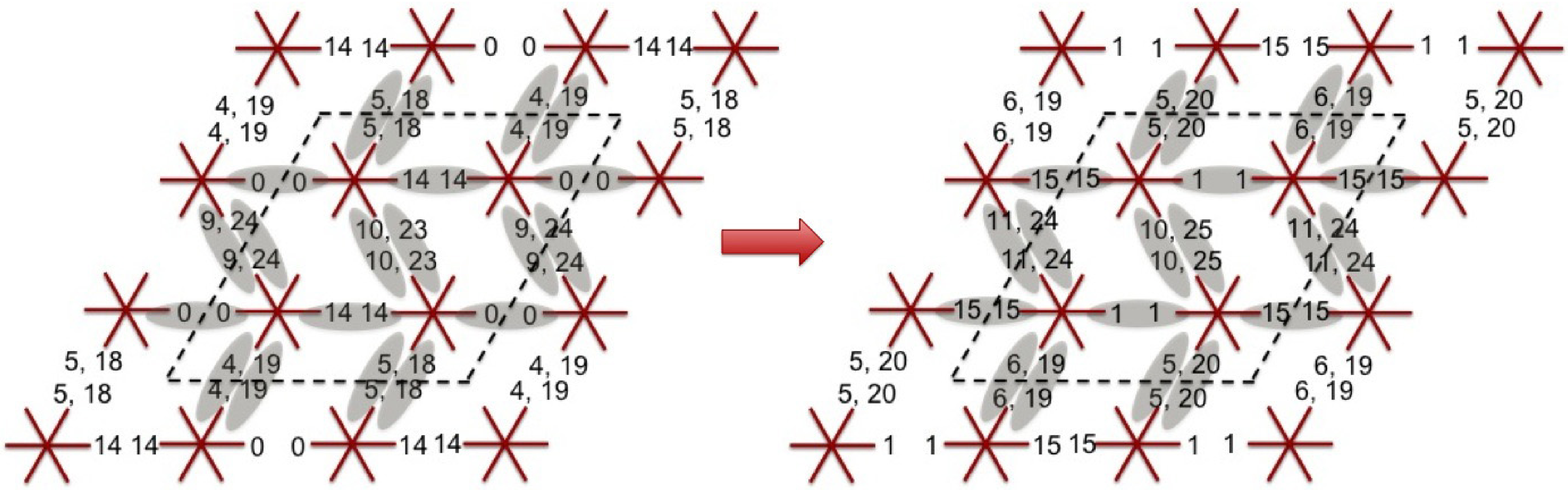} 
\caption{The crosslinking structure in the right is generated by the $2\pi/28$ coherent rotation of maximally crosslinked bundle groundstate shown in the left, followed by a reorganization of monomers and crosslinks within a 28-monomer repeat length.  Note that under rotation the layers where crosslinkings occur shift, while the crosslinking directions are  preserved. } 
\label{fig: rotation}
\end{figure}

\begin{table}
\begin{tabular}{|c||c|c|c|c|c|c|}
\hline
\multirow{2}{*}{$n_b=0$} &$s_0$ & $\mid28\mid$ &   &   & &   \\ \cline{2-7}
  &$\mu_c$ & $-\epsilon_b$&  &  & & \\ \hline \hline
\multirow{2}{*}{$n_b=1$} &$s_1$ & $\mid28\mid$ &   &   & &   \\ \cline{2-7}
  &$\mu_c$ & $-\epsilon_b$&  &  & &\\  \hline \hline
  \multirow{2}{*}{$n_b=2$} &$s_2$ & $\mid14\mid14\mid$ & $\mid 5 \mid 23 \mid$  & $\mid 8 \mid20\mid$  & $\mid 10 \mid 18\mid$ & $\mid 13\mid15\mid$  \\ \cline{2-7}
  &$\mu_c$ & $-\epsilon_b$
  & $-\epsilon_b+0.00068C$ & $-\epsilon_b+0.00783C$ & $-\epsilon_b+0.00174C$ &$-\epsilon_b+0.00361C$ \\ \hline \hline
\multirow{2}{*}{$n_b=3$}  & $s_3$ &  $\mid 4 \mid 10 \mid 14\mid $  & $\mid 9 \mid 5 \mid 14\mid$  & $\mid 13 \mid 5 \mid 10\mid$  & $\mid 8 \mid5 \mid 15 \mid$  &   \\\cline{2-7}
  &$\mu_c$ & $-\epsilon_b+0.00261C$ & $-\epsilon_b+0.00058C$ &  $-\epsilon_b+0.00241C$ &$-\epsilon_b+ 0.00522C$ & \\ \hline \hline
\multirow{2}{*}{$n_b=4$}  & $s_4$ &  $\mid 4 \mid 5 \mid 5\mid 14\mid $ & $\mid 4 \mid 10 \mid 5\mid 9\mid $  & $ \mid 4\mid 10 \mid 10\mid 4\mid$ & $\mid 9 \mid 5 \mid 5\mid 9\mid$ & $\mid 8 \mid 5 \mid 10 \mid 5\mid$  \\ \cline{2-7}
   &$\mu_c$ & $-\epsilon_b+0.00195C$ & $-\epsilon_b+0.00239C$ & $-\epsilon_b+0.00391C$ & $-\epsilon_b+0.00087C$ & $-\epsilon_b+0.00391C$ \\ \hline \hline
\multirow{2}{*}{$n_b=5$}  &$s_5$ & $\mid 4 \mid 5\mid 5\mid 5 \mid 9\mid $ & $\mid 4 \mid 5 \mid 5 \mid 10 \mid 4\mid$ & & &   \\ \cline{2-7}
 &$\mu_c$ & $-\epsilon_b+0.00191C$ &  $-\epsilon_b+0.00313C$ & & & \\ \hline \hline
\multirow{2}{*}{$n_b=6$}  &$s_6$ & $\mid 4 \mid 5 \mid 5 \mid 5\mid 5\mid 4\mid$ & & & &   \\ \cline{2-7}
&$\mu_c$ &$-\epsilon_b+0.00261C$ & & & & \\ \hline 
\end{tabular}
\caption{The crosslinking configurations commensurate with the -28/13 filament geometry for all $n_b$ and the corresponding elastic offset energy, $\mu_c(n_b,s_{n_b})$.  }
\label{tab:table}
\end{table}

\begin{figure}
\centering
\includegraphics[angle=-90.0, scale=0.5]{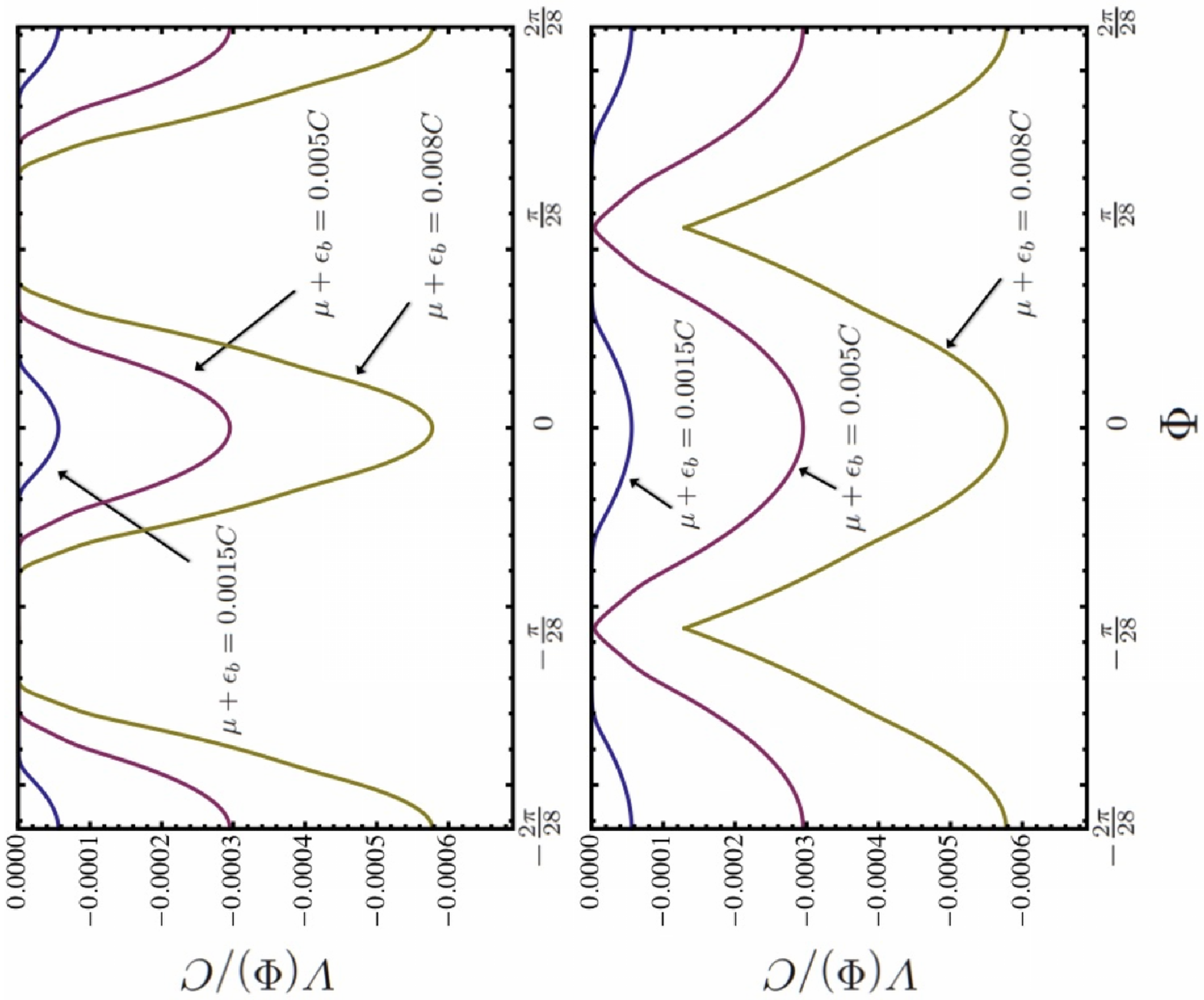} 
\caption{The $\Phi$ dependence of the crosslinking free energy at $\beta C=6900$ for $U=3.0C$ (top) and $U=0.5C$ (bottom) for various chemical potentials.  Finite temperature effects from the statistical distribution of crosslinks in the bundle lead to modest quantitative differences with Fig. 3 of the text. } 
\label{fig: potential_finiteT}
\end{figure}

\begin{figure}
\centering
\includegraphics[angle=0.0, scale=0.5]{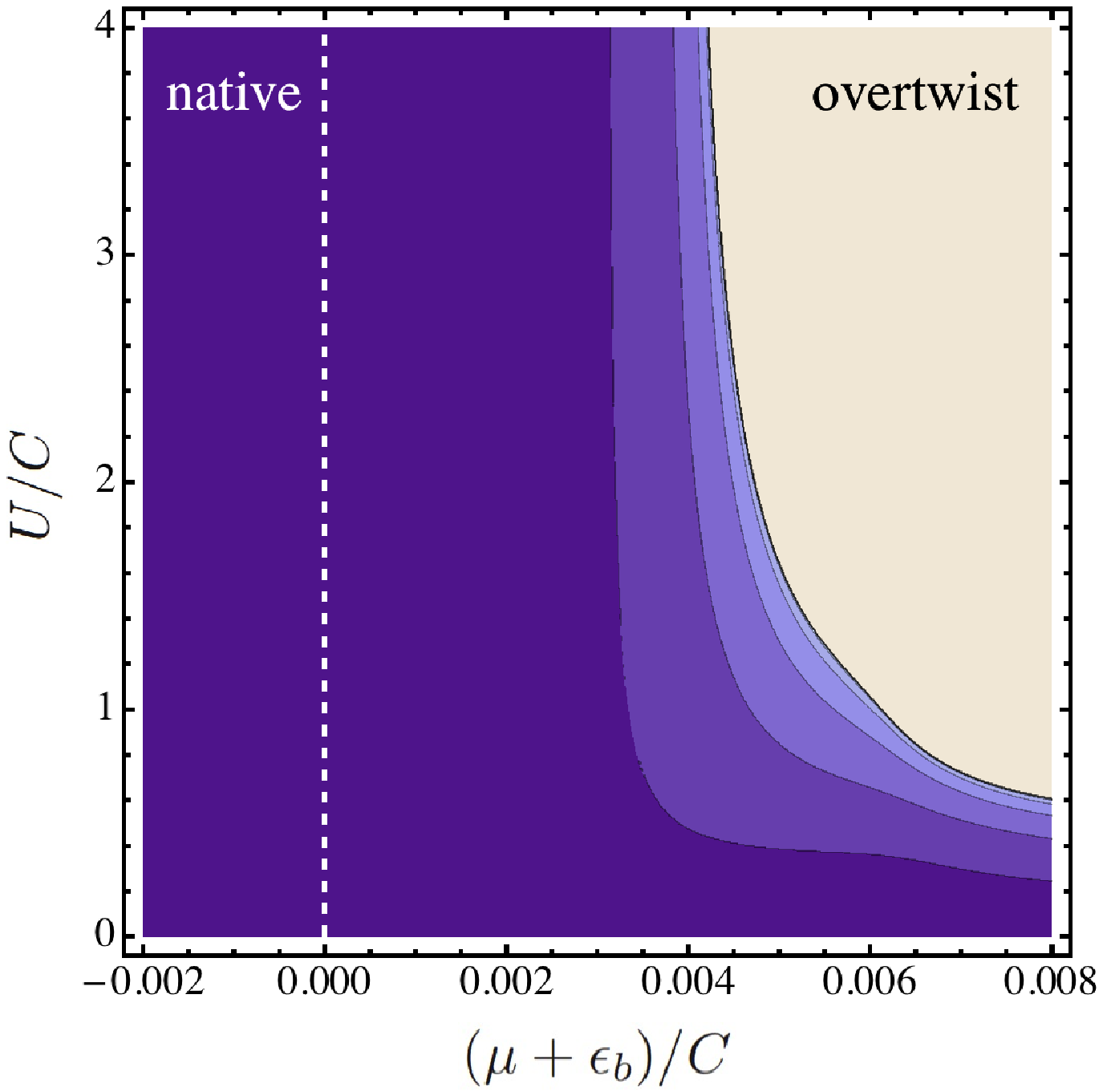} 
\caption{The equilibrium diagram of state for overtwist, $\langle \Delta \phi \rangle$, for $\beta C =6900$.  The CI transition is shown as a solid line.   } 
\label{fig: finiteT}
\end{figure}

\begin{figure}
\centering
\includegraphics[angle=0.0, scale=0.5]{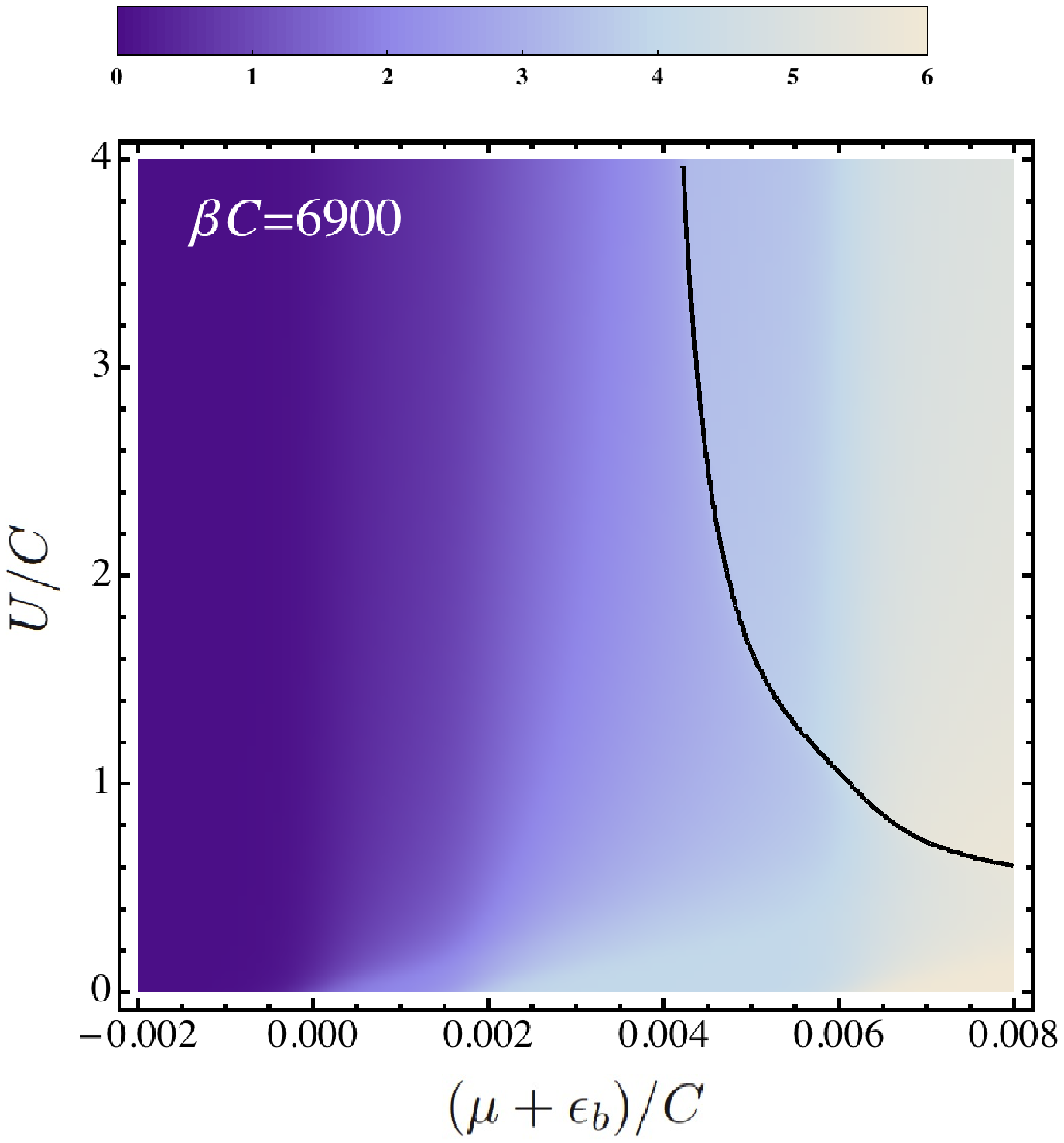} 
\caption{The mean number of crosslinker per 28-monomer repeat, $\langle n_b \rangle$, for $\beta C=6900$.   The solid line denotes the CI transtion.} 
\label{fig: bonddensity_finiteT}
\end{figure}

\end{document}